\documentclass{aa}

\usepackage{graphicx}
\usepackage{amsmath}
\usepackage{amssymb}
\usepackage{natbib}
\usepackage{hyperref}
\usepackage{xcolor}
\usepackage{booktabs}
\usepackage{multirow}

\begin{document}

\title{A statistically robust framework for detecting and classifying hysteresis patterns in astrophysical spectral evolution}
\titlerunning{Hysteresis detection in astrophysics}

\author{T. Terzi\'{c}\inst{1}\fnmsep\thanks{Corresponding author: tterzic@phy.uniri.hr}}

\institute{University of Rijeka, Faculty of Physics, 51000 Rijeka, Croatia}

\date{Received \today; accepted --}

\abstract
{Loop-like patterns between spectral parameters are frequently interpreted as evidence of hysteresis in time-dependent astrophysical emission processes. Such patterns have been reported in hardness-intensity diagrams (HID) of accreting black-hole X-ray binaries during state transitions, in the radio-to-X-ray correlation plane during outbursts, in solar activity indices over the solar cycle, and in the spectral energy distribution of active galactic nuclei during flaring episodes. In the last case, HID or the evolution of the synchrotron peak frequency with the corresponding flux often exhibit apparent clockwise or counter-clockwise loops, whose orientation encodes the relative timescales of particle acceleration and radiative cooling. Visual inspection, however, does not provide a statistically controlled detection method, and the false positive rate associated with stochastic variability remains poorly quantified.}
{We develop a statistically robust and empirically calibrated framework for detecting, quantifying, and classifying hysteresis patterns in ordered two-dimensional data with measurement uncertainties.}
{The method is based on the signed geometric area enclosed by chronologically ordered points in the plane, computed using the \textit{shoelace formula}. We define open and closed area estimators, introduce cancellation diagnostics for multi-loop structures, and propagate measurement uncertainties via Monte Carlo (MC) sampling. Statistical significance is assessed using null ensembles generated by time-order randomization, physically motivated autoregressive surrogate models, and Fourier phase-randomized surrogates, providing empirical $p$-values against each null hypothesis.}
{The framework provides the normalised signed area $A_\mathrm{norm}$ as the primary detection statistic, complementary shape diagnostics, MC uncertainty intervals, and empirical $p$-values against three null models. We validate the method on synthetic blazar flare trajectories representative of high-synchrotron-peaked BL Lac objects, and demonstrate its application to an XMM-Newton observation of Markarian~421 during a December 2023 flaring episode, where we confirm a CCW hysteresis loop with $A_\mathrm{norm} = +0.64$ that is robust against measurement noise but does not reach formal significance against stochastic null models, possibly due to the open trajectory geometry. We provide fully documented open-source Python software for community use.}
{}

\maketitle
\nolinenumbers

%%%%%%%%%%%%%%%%%%%%%%%%%%
\section{Introduction}
\label{sec:intro}

Time-dependent astrophysical emission processes frequently exhibit complex spectral evolution during transient or flaring episodes. Loop-like trajectories in two-dimensional diagnostic planes are commonly interpreted as signatures of hysteresis in time-variable astrophysical objects. Such patterns arise when two observables respond to a common driver with different temporal delays or nonlinear couplings, producing oriented loops when plotted against each other in chronological order.

Hysteresis behaviour has been reported in several astrophysical contexts. 
In accreting black-hole X-ray binaries (BHXBs), two distinct manifestations have been studied. The first concerns the spectral state transition itself: hardness-intensity diagrams (HIDs) exhibit clockwise (CW) and counter-clockwise (CCW) loops during the rise and decay of outbursts, because the source follows different paths between the soft and hard states depending on the direction of evolution \citep{miyamoto1995,zdziarski2004,belloni2005,2025MNRAS.541.1851B}. This HID hysteresis reflects the dependence of the accretion geometry and jet activity on the outburst history, not merely on the instantaneous luminosity \citep{dunn2010}. 
The second manifestation concerns the radio-to-X-ray correlation: the same BHXBs trace distinct tracks during the rise and decay of their outbursts, producing loop-like deviations from the mean correlation in the radio/X-ray plane \citep{corbel2013}. Although both effects occur in the same class of sources and often during the same outburst, they operate in different diagnostic planes and reflect different physical couplings---the HID hysteresis probes the disk-corona transition, while the radio/X-ray hysteresis probes jet formation and quenching.

In blazars, the HID or the synchrotron peak frequency and corresponding peak flux often trace loop-like trajectories during flaring episodes \citep{1996ApJ...470L..89T, 2000ApJ...541..166F, 2017ApJ...834....2A, MAGIC:2024mjz, MAGIC:2025omm}. These loops arise from the different timescales of particle acceleration and synchrotron cooling: when the acceleration timescale is shorter than the cooling timescale, the peak frequency rises before the flux, producing CW loops; the opposite produces CCW loops \citep{tramacere2009}. 

Solar activity indices display analogous hysteresis effects over the solar cycle when magnetic and radiative proxies are compared \citep{1994SoPh..150..347B,2003NewA....8..745O,2012SoPh..276..407S}.

Despite widespread qualitative identification of such loops, there is no generally adopted statistical framework for quantifying their significance. Visual inspection cannot distinguish genuine phase-lagged physical evolution from stochastic variability, measurement noise, sparse sampling, or red-noise processes (i.e.\ stochastic variability with power spectral density $P(f) \propto f^{-\beta}$, $\beta > 0$), which are ubiquitous in accreting systems and blazars \citep{uttley2002,vaughan2003a,mchardy2004,emmanoulopoulos2013}.
The absence of calibrated false positive rates makes it difficult to assess whether observed loop areas are statistically significant. Furthermore, many trajectories are not geometrically closed; forcing closure can introduce artificial area contributions that mimic hysteresis.

In this work, we develop a statistically controlled framework to detect, quantify, and classify hysteresis patterns in ordered two-dimensional datasets with measurement uncertainties in both variables. 
The method uses the signed geometric area of the chronological trajectory as the primary statistic. It distinguishes open-path area from artificial closure contributions, introduces cancellation diagnostics for multi-loop structures, propagates measurement uncertainties via Monte Carlo (MC) sampling, constructs empirically calibrated null ensembles, and provides $p$-values for statistical tests. 

The paper is structured as follows. Sect.~\ref{sec:geometry} defines the geometric framework, uncertainty propagation and null models are described in Sect.~\ref{sec:uncertainty}, and the method is validated on synthetic trajectories in Sect.~\ref{sec:visualisation}. The decision framework is presented in Sect.~\ref{sec:decision}. The method is applied on real data in Sect.~\ref{sec:application}. Our work is summarised in Sect.~\ref{sec:conclusions}. 
The accompanying open-source Python software is available at \url{https://github.com/tterzic/hysteresis-detection}.

%%%%%%%%%%%%%%%%%%%%%%%%%%
\section{Geometric framework}
\label{sec:geometry}

The defining feature of hysteresis is that the trajectory traced in observable space depends on the temporal ordering of the measurements. Forward and backward evolution need not follow the same path, leading to loop-like structures when consecutive points are connected in time order.

We consider a sequence of $N$ measurements obtained at times $t_1 < t_2 < \cdots < t_N$, each represented by a point
\begin{equation}
  (x_i, y_i) = (x(t_i),\, y(t_i)),
  \label{eq:points}
\end{equation}
with associated measurement uncertainties $(\sigma_{x,i}, \sigma_{y,i})$ that may vary between individual measurements.

\subsection{Loop area as a measure of hysteresis}
\label{sec:looparea}

We associate a signed geometric area with the trajectory by connecting consecutive measurements with straight line segments. The signed area of the resulting closed polygon is given by the familiar shoelace formula,
\begin{equation}
  A_\mathrm{tot} = \sum_{i=1}^{N} a_i, \qquad
  a_i = \frac{1}{2}(x_i y_{i+1} - x_{i+1} y_i),
  \label{eq:atot}
\end{equation}
with $(x_{N+1}, y_{N+1}) \equiv (x_1, y_1)$. The magnitude of $A_\mathrm{tot}$ quantifies the strength of hysteresis, while its sign encodes the direction of traversal: positive $A_\mathrm{tot}$ implies CCW, and negative indicates CW direction under the right-hand rule). The area is invariant under translations and rotations of the coordinate system, and it is zero when the forward and backward paths coincide.

\subsection{Open and closed area decomposition}
\label{sec:openclosed}

The last point in the sequence typically does not coincide with the first, since observations often do not follow the source through a complete cycle. Enforcing geometric closure by connecting $(x_N, y_N)$ back to $(x_1, y_1)$ introduces an additional contribution. We therefore separate the area as
\begin{equation}
  A_\mathrm{tot} = A_\mathrm{open} + A_\mathrm{closure},
  \label{eq:decomp}
\end{equation}
where $A_\mathrm{open} = \sum_{i=1}^{N-1} a_i$ is the area accumulated along the time-ordered segments, and $A_\mathrm{closure} = \frac{1}{2}(x_N y_1 - x_1 y_N)$ is the contribution of the artificial closing segment. 
Since open loops are more common in real datasets, $A_\mathrm{open}$ is the primary statistic throughout this work; $A_\mathrm{tot}$ is never used as the detection statistic.

$A_\mathrm{open}$ and $A_\mathrm{closure}$ are not translation- and rotation-invariant, which prevents comparison between datasets. Fortunately, this can be easily fixed by choosing the origin of the coordinate system in the centroid of the dataset. 
Therefore, we first translate all coordinates to the centroid of the $N$ observed points,
\begin{equation}
\begin{aligned}
  \tilde{x}_i = x_i - \bar{x}, \quad \bar{x} = \frac{1}{N}\sum_{i=1}^N x_i, \\
  \tilde{y}_i = y_i - \bar{y}, \quad \bar{y} = \frac{1}{N}\sum_{i=1}^N y_i,
\end{aligned}
\label{eq:centroid}
\end{equation}
and work with the centred coordinates $(\tilde{x}_i, \tilde{y}_i)$ throughout. 
Now, the incremental signed triangle areas are
\begin{equation}
  a_i = \frac{1}{2}(\tilde{x}_i\,\tilde{y}_{i+1} -
                    \tilde{x}_{i+1}\,\tilde{y}_i), \quad A_\mathrm{closure} = a_N.
  \label{eq:ai}
\end{equation}
All previously given expressions remain the same. 
For the sake of simplicity, from here on, we will use $(x_i, y_i)$ for the centred coordinates. 

To warn when the forced closure segment makes a large contribution, we define the closure fraction
\begin{equation}
  f_\mathrm{cl} = \frac{|A_\mathrm{closure}|}{|A_\mathrm{tot}|}.
  \label{eq:fclosure}
\end{equation}
The values $f_\mathrm{cl} \ll 1$ indicate a negligible contribution from endpoint linking; $f_\mathrm{cl} \sim 1$ indicates that most of the area arises from forced closure. This is a diagnostic flag only, the quantity being analysed is always $A_\mathrm{open}$. 

\subsection{Normalized closure distance}
\label{sec:closuredist}

We also define a normalised closure distance to indicate whether the trajectory endpoints are statistically consistent with closure: 
\begin{equation}
  d_\mathrm{cl} = \sqrt{
    \frac{(x_N - x_1)^2}{\sigma_{x,N}^2 + \sigma_{x,1}^2} +
    \frac{(y_N - y_1)^2}{\sigma_{y,N}^2 + \sigma_{y,1}^2}}.
  \label{eq:dclosure}
\end{equation}
This measures the endpoint separation in units of the combined measurement uncertainties and serves as a qualitative guide to determining whether the trajectory is consistent with the closure. Small values indicate consistency with closure and large values indicate a significant geometric opening. 
For $d_\mathrm{cl} \lesssim 0.5$, the closure contribution $|A_\mathrm{closure}|$ is typically less than 10\% of $|A_\mathrm{open}|$ (median value across simulated flare trajectories with $N = 10$--$40$), and $A_\mathrm{open}$ and $A_\mathrm{tot}$ are statistically indistinguishable for most practical purposes. For $d_\mathrm{cl} \gtrsim 2$, the closure contribution is comparable to or greater than $|A_\mathrm{open}|$ for a substantial fraction of trajectories.
Large values of $d_\mathrm{cl}$ indicate that the source was not observed through a complete cycle, but do not imply by themselves that $A_\mathrm{open}$ is not a valid measurement. 

The closure fraction $f_\mathrm{cl}$ and the normalised closure distance $d_\mathrm{cl}$ are largely non-redundant. Their relationship is discussed in Appendix~\ref{app:closure}. 

The raw endpoint separation 
\begin{equation}
    \Delta_\mathrm{obs} = \sqrt{(x_N - x_1)^2 + (y_N - y_1)^2} 
    \label{eq:Delta_obs}
\end{equation}
and its propagated 1-sigma uncertainty
\begin{equation}
  \sigma_\Delta = \frac{\sqrt{(x_N-x_1)^2(\sigma_{x,N}^2+\sigma_{x,1}^2) + (y_N-y_1)^2(\sigma_{y,N}^2+\sigma_{y,1}^2)}}{\Delta_\mathrm{obs}}
  \label{eq:sigmadelta}
\end{equation}
are related to but not equivalent to $d_\mathrm{cl}$. While both $d_\mathrm{cl}$ and the ratio $\Delta_\mathrm{obs}/\sigma_\Delta$ measure the significance of the endpoint separation relative to the measurement uncertainties, they weight the contributions of $x$ and $y$ differently: $d_\mathrm{cl}$ normalises the displacement in each coordinate separately by the combined uncertainty in that coordinate, whereas $\sigma_\Delta$ propagates the uncertainties through the Euclidean distance. The two quantities coincide only when the uncertainties at both endpoints are equal in each coordinate. We report $d_\mathrm{cl}$ rather than $\Delta_\mathrm{obs}$ and $\sigma_\Delta$ separately because it is a single dimensionless number summarising both the magnitude of the separation and its significance relative to the measurement uncertainties. 

\subsection{Normalized area}
\label{sec:normedarea}

To allow scale-invariant comparison between datasets, we normalise $A_\mathrm{open}$ by the area of the convex hull of the centred data points $(x_i, y_i)$:
\begin{equation}
  A_\mathrm{norm} = \frac{A_\mathrm{open}}{A_\mathrm{hull}},
  \label{eq:anorm}
\end{equation}
where $A_\mathrm{hull}$ is the convex hull area. 

The convex hull of a set of points is the smallest convex polygon that contains all the points. Geometrically, it is the shape obtained by stretching a rubber band around the outermost points. Its area is computed by applying the shoelace formula to the $M \leq N$ hull vertices listed in order,
\begin{equation}
  A_\mathrm{hull} = \frac{1}{2} \left|
  \sum_{k=1}^{M} (x_{v_k}\,y_{v_{k+1}} - x_{v_{k+1}}\,y_{v_k})
  \right|,
  \label{eq:ahull}
\end{equation}
with $(v_{M+1}) \equiv (v_1)$, where $v_1, \ldots, v_M$ are the indices of the hull vertices in cyclic order. Since the convex hull is always a simple closed polygon, the absolute value ensures $A_\mathrm{hull} > 0$ regardless of direction and chronological order of the vertices. $A_\mathrm{hull}$ depends only on the spatial distribution of the $N$ points and not on their chronological order, making it a natural normalisation factor that captures the spatial extent of the data independently of the shape of the trajectory. 

Defined in this way, $A_\mathrm{norm}$ is translation- and rotation-invariant.
For simple (non-self-intersecting) trajectories, $|A_\mathrm{norm}| \leq 1$ by construction, since the open-path shoelace area of a simple polygonal path cannot exceed the area of the convex hull of its vertices. Self-intersecting trajectories that wind multiple times in the \emph{same} direction are an exception: the shoelace formula assigns a winding number to each enclosed region, so a path that winds twice around the same region counts that region's area twice, yielding $|A_\mathrm{norm}| > 1$. In practice such cases are rare (see Appendix~\ref{app:winding} for a quantitative characterisation across all three null models) and have negligible practical effect on the reported $p$-values.

If all points are collinear, $A_\mathrm{hull} = 0$ and every $a_i = 0$, so $A_\mathrm{open} = 0$; in this case $A_\mathrm{norm}$ is undefined and the data contain no loop structure. 

All statistical calibrations are performed on $A_\mathrm{norm}$.

\subsection{Cancellation diagnostics}
\label{sec:cancel}

Multi-loop or figure-eight structures may produce a small net $A_\mathrm{open}$ due to cancellation between segments rotating in opposite directions. We therefore define the absolute incremental area
\begin{equation}
  A_\mathrm{abs} = \sum_{i=1}^{N-1} |a_i|
  \label{eq:aabs}
\end{equation}
and the root-mean-square area
\begin{equation}
  A_\mathrm{rms} = \left(\sum_{i=1}^{N-1} a_i^2\right)^{1/2}.
  \label{eq:arms}
\end{equation}
$A_\mathrm{abs}$ is adopted as the primary magnitude diagnostic (see Appendix~\ref{app:arms} for a detailed justification and
comparison with $A_\mathrm{rms}$). The cancellation ratio 
\begin{equation}
  R_\mathrm{can} = \frac{|A_\mathrm{open}|}{A_\mathrm{abs}}
  \label{eq:rcancel}
\end{equation}
has $R_\mathrm{can} \approx 1$ for coherent single-loop trajectories and $R_\mathrm{can} \ll 1$ for strongly cancelling
multi-loop structures.

%%%%%%%%%%%%%%%%%%%%%%%%%%
\section{Uncertainty estimation and null models}
\label{sec:uncertainty}

\subsection{Propagation of measurement uncertainties}
\label{sec:mcprop}

The signed area estimator is nonlinear in the data and depends on products of adjacent measurements. Therefore, we propagate measurement uncertainties via MC sampling. 
For each realisation $k$, synthetic measurements are generated as
\begin{equation}
  x_i^{(k)} \sim \mathcal{N}(x_i, \sigma_{x,i}), \quad
  y_i^{(k)} \sim \mathcal{N}(y_i, \sigma_{y,i}),
  \label{eq:mcsampling}
\end{equation}
and $A_\mathrm{norm}$ is recomputed for each realisation. 
We use $K_\mathrm{MC} = 10^4$ realisations for uncertainty estimation. 
The ensemble $\{A^{(k)}\}$ provides the MC mean $\langle A \rangle$, standard deviation $\sigma_\mathrm{MC}$ and confidence intervals (CI) from empirical quantiles, quantifying the robustness of the inferred orientation. This distribution is not a null hypothesis for hysteresis, it quantifies the sensitivity of the measured area to observational noise.

\subsection{Stochastic variability mimicking hysteresis}
\label{sec:null_motivation}

A non-zero $A_\mathrm{open}$ does not by itself imply genuine hysteresis. AGN and X-ray binaries are well known to exhibit red-noise variability: their flux fluctuations are temporally correlated, with power spectral density $P(f) \propto f^{-\beta}$ \citep{uttley2002,Paolillo:2023ssv}. 
Red-noise processes can generate consecutive observations trending in the same direction for extended periods. When two such correlated but physically unrelated time series are plotted against each other in chronological order, the resulting trajectory can trace an apparent loop. Without null models that account for this autocorrelation structure, such stochastic loops can easily be mistaken for physical hysteresis. 
To address this, we compare $A_\mathrm{norm}$ of the observed data with distributions from three classes of simulated null realisations (hereafter also referred to as surrogates).

\subsubsection{Null hypothesis I: random time ordering}
\label{sec:nullperm}

The simplest null hypothesis asks whether the observed loop could arise from a different temporal ordering of the same observed data points. We construct simulated null realisations by randomly permuting the time ordering of the pairs $(x_i, y_i)$, keeping each pair intact while shuffling their sequence in time. This preserves the marginal distributions of both variables and their mutual correlation within each pair, while destroying the temporal evolution. The two-sided empirical $p$-value is
\begin{equation}
  p_\mathrm{perm} = \frac{1}{K} \sum_{k=1}^{K}
  \mathbf{1}\!\left(|A_\mathrm{norm}^{(\pi_k)}| \geq |A_\mathrm{norm,obs}|\right),
  \label{eq:pvalue}
\end{equation}
where $\pi_k$ denotes the $k$-th random permutation and $\mathbf{1}$ is the indicator function (equal to 1 if its argument is true and 0 otherwise). 
This $p$-value requires no distributional assumptions \citep{davison1997}. 

We adopt a two-sided test, using $|A_\mathrm{norm}^{(\pi_k)}| \geq |A_\mathrm{norm,obs}|$, because in most astrophysical scenarios the orientation of the loop is not known a priori. Therefore, the primary detection criterion tests for the presence of hysteresis of either sign, while orientation is reported separately via the sign of $A_\mathrm{norm}$. 

\subsubsection{Null hypothesis II: correlated stochastic variability}
\label{sec:arsurr}

Astrophysical light curves frequently exhibit strong temporal autocorrelation \citep{vaughan2003a,uttley2005}: a high flux at one epoch makes a high flux at the next more likely.\footnote{This is a direct consequence of the red-noise power spectrum. By the Wiener--Khinchin theorem, the autocorrelation function $R(\tau) = \langle x(t)\, x(t+\tau)\rangle$ is the inverse Fourier transform of the power spectral density (PSD). For $P(f) \propto f^{-\beta}$ with $\beta > 0$, power is concentrated at low frequencies, corresponding to slow variations that persist over many consecutive time steps. Consequently, $R(\tau) > 0$ over a broad range of lags, meaning a high flux at one epoch tends to be followed by high flux at subsequent epochs.}
Shuffling time pairs alone underestimates the probability of spurious loops because the shuffled surrogates lack the continuous drifting behaviour of red-noise variability. 

We therefore construct an additional null using independent first-order autoregressive AR(1) processes \citep{brockwell2002}:
\begin{equation}
\begin{aligned}
  x_i = \phi_x(x_{i-1} - \mu_x) + \mu_x + \epsilon_i,\\
  y_i = \phi_y(y_{i-1} - \mu_y) + \mu_y + \eta_i,
\end{aligned}
\label{eq:ar1}
\end{equation}
where $\epsilon_i$ and $\eta_i$ are independent Gaussian random variables introducing the unpredictable increment added at each time step.  
$\epsilon_i \sim \mathcal{N}(0, \sigma_x^2)$ and $\eta_i \sim \mathcal{N}(0, \sigma_y^2)$ are independent, with
$\sigma_x^2 = \mathrm{Var}(x)(1 - \hat{\phi}_x^2)$ and
$\sigma_y^2 = \mathrm{Var}(y)(1 - \hat{\phi}_y^2)$.
The autoregressive coefficient $\phi$ is estimated from the observed time series using the Yule--Walker method of moments \citep{brockwell2002}: $\hat{\phi} = \sum_{i=1}^{N-1} x_i x_{i+1} / \sum_{i=1}^{N-1} x_i^2$, and analogously for $y$. 
The variances of $\epsilon_i$ and $\eta_i$ ensure that the simulated series reproduces the observed variance. 
The autoregressive coefficient captures the short-timescale correlation structure of the data to first order. Higher-order correlations are not explicitly matched, and the AR(1) model should be regarded as a conservative but approximate representation of the red-noise variability present in real astrophysical light curves. The two series are generated independently, so there is no physical coupling between them. 

For datasets with more complex autocorrelation structures, the AR(1) model can be replaced by an AR($p$) process selected using the Akaike information criterion \citep{akaike1974}.

\subsubsection{Null hypothesis III: Fourier phase randomisation}
\label{sec:fouriersurr}

A third class of null models preserves the autocorrelation structure of each observed time series while removing any coherent phase relationship between the two variables. Using the Timmer~\&~K\"{o}nig algorithm \citep{timmer1995}, the Fourier amplitudes of each series are preserved and the phases are replaced by independent uniform random draws from $[0, 2\pi\rangle$. The inverse Fourier transform yields surrogates with the same power spectrum---and therefore the same autocorrelation structure---as the original data, but with no cross-variable coherence. This null is distinct from the AR(1) null in that it makes no parametric assumption about the form of the autocorrelation; instead it uses the observed power spectrum directly. 
The Fourier null distribution of $A_\mathrm{norm}$ is bimodal for some trajectories and unimodal for others, depending on the spectral content of the specific dataset. Bimodality arises when the power spectrum is dominated by a single low-frequency component: the phase of that mode largely determines the direction of the resulting loop, so with phases drawn uniformly from $[0, 2\pi\rangle$, roughly half the surrogates produce a CW loop and half a CCW one of similar area, giving two peaks near $\pm|A_\mathrm{norm}|$. In simulations using $K = 10^4$ Fourier surrogates for ten trajectories drawn from the toy model at each $N$ between 6 and 20, we find that bimodality strength varies substantially from trajectory to trajectory at fixed $N$, with no systematic trend in $N$. The two-sided $p$-value used here compares $|A_\mathrm{norm}^\mathrm{obs}|$ against $|A_\mathrm{norm}^\mathrm{null}|$. For a symmetric bimodal distribution this is equivalent to a one-sided test restricted to surrogates of the same orientation as the observed loop. 

The Fourier null, as implemented, operates on the sequence of observed values, treating the index as a uniform time axis. This is a valid characterisation of the autocorrelation structure for roughly evenly sampled data, but it may not accurately represent the underlying process for strongly irregular sampling, in which case $p_\mathrm{Fourier}$ should be interpreted with additional caution.
The Fourier null requires at least $N = 6$ observations. For shorter series, the discrete Fourier transform has too few independent frequency bins to produce a meaningful surrogate distribution, and the method raises an error. For small $N$ generally, the observed power spectrum is a noisy estimate of the underlying process, and $p_\mathrm{Fourier}$ should be interpreted with correspondingly greater caution.

\subsubsection{Combined null ensemble}
\label{sec:combined}

For a comprehensive assessment, we define the combined null ensemble
\begin{equation}
  \mathcal{N}_\mathrm{full} = \mathcal{N}_\mathrm{perm} \cup \mathcal{N}_\mathrm{Fourier} \cup \mathcal{N}_\mathrm{AR},
  \label{eq:combined}
\end{equation}
where $\mathcal{N}_\mathrm{perm}$, $\mathcal{N}_\mathrm{Fourier}$, and $\mathcal{N}_\mathrm{AR}$ are the sets of null loop areas under the permutation, AR(1), and Fourier nulls. The combined $p$-value is
\begin{equation}
  p_\mathrm{full} = \frac{
    \#\!\left\{ A^{(k)} \in \mathcal{N}_\mathrm{full} : |A^{(k)}| \geq |A_\mathrm{obs}| \right\}}
    {|\mathcal{N}_\mathrm{full}|},
  \label{eq:pfull}
\end{equation}
where $A^{(k)}$ is the normalised loop area of the $k$-th null realisation (from any of the three null classes) and $|\mathcal{N}_\mathrm{full}|$ is the total number of surrogates pooled across all three nulls. The fraction counts how many surrogates exceed the observed area. This provides a conservative summary $p$-value across all considered stochastic mechanisms. 

Because each null model contributes equally to $\mathcal{N}_\mathrm{full}$, this is equivalent to evaluating the observed area against a mixture distribution
\begin{equation}
  p_\mathrm{full}(a) = \tfrac{1}{3}\,p_\mathrm{perm}(a)
                     + \tfrac{1}{3}\,p_\mathrm{AR}(a)
                     + \tfrac{1}{3}\,p_\mathrm{Fourier}(a),
  \label{eq:pfull_mixture}
\end{equation}
which implicitly assigns equal prior probability to each of the three null hypotheses. This prior is a pragmatic choice in the absence of external knowledge about which stochastic mechanism is most plausible. It is not formally optimal but is transparent and easy to update if one null model is deemed more appropriate for a given dataset.
The value of $p_\mathrm{full}$ is not guaranteed to be the most conservative of the four $p$-values. Indeed, it lies between the most conservative and least conservative individual values, weighted equally across the three nulls. Its principal virtue is robustness to the choice of null model, i.e. small $p_\mathrm{full}$ means that the observed loop is difficult to explain under any of the three null hypotheses simultaneously. 

We adopt $p_\mathrm{full}$ as the primary significance statistic throughout this work. 
However, we recommend using all three null models and reporting all three individual $p$-values ($p_\mathrm{perm}$, $p_\mathrm{AR}$, $p_\mathrm{Fourier}$) alongside $p_\mathrm{full}$ from the combined null ensemble. The individual $p$-values carry distinct physical meaning: a small $p_\mathrm{perm}$ indicates that the loop is unlikely to arise from a different temporal ordering of the same data; a small $p_\mathrm{AR}$ indicates that it is unlikely to arise from correlated red-noise variability alone; and a small $p_\mathrm{Fourier}$ indicates that it is unlikely to arise from a stochastic process with the same power spectrum. A researcher may legitimately report a statistically significant loop under the permutation null while acknowledging that the AR(1) or Fourier nulls cannot be rejected, which would support the presence of a coherent loop structure but stop short of claiming a specific physical mechanism as its origin.

\subsection{Open trajectories and detection power}
\label{sec:opentrajectories}

Trajectories that do not complete a full cycle (flagged by large $f_\mathrm{cl}$ or $d_\mathrm{cl}$) have lower $|A_\mathrm{norm}|$ than comparable closed loops, because the open-path area $A_\mathrm{open}$ does not include the contribution that the closing segment would have provided. This reduces detection power because an open observed trajectory must produce a larger true loop to reach the same $|A_\mathrm{norm}|$ as a closed one. 
Therefore, $f_\mathrm{cl}$ and $d_\mathrm{cl}$ should be treated as qualitative indicators of reduced detection power rather than as corrections to the reported $p$-values.

We investigated applying two possible corrections to $|A_\mathrm{norm}|$ in the case of open loops in Appendix~\ref{app:corrections}. However, we identified serious objections to both of these procedures. 
We therefore retain $|A_\mathrm{norm}|$ as the primary test statistic and do not apply the closure-fraction correction.

%%%%%%%%%%%%%%%%%%%%%%%%%%
\section{Visualisation and calibration}
\label{sec:visualisation}

We validate the framework using simulations of physically motivated hysteresis trajectories. We focus on blazar spectral evolution as the primary example, although a similar exercise can be performed on any physical system showing loop-like behaviour. 

\subsection{Synthetic blazar flare model}
\label{sec:model}

Although the method developed in this paper is entirely model-independent and can be applied to any pair of
simultaneously measured observables, the validation simulations require a concrete choice of diagnostic plane. We adopt the hardness ratio--flux ($HR$--$F$) plane, where $HR$ is the ratio of hard-band to soft-band flux and $F$ is the total X-ray flux. This plane is the standard diagnostic for blazar spectral evolution in X-ray observations \citep{tramacere2009, Kapanadze:2020qpy, Kapanadze:2024nwg, MAGIC:2024mjz, MAGIC:2025omm} and has been used explicitly to study hysteresis patterns and their physical interpretation in terms of particle acceleration and cooling \citep[e.g.][]{kirk1998,fossati2000a,Levy:2024eiq}. 
%For HBL sources such as Markarian~421, the hardness ratio tracks the position of the synchrotron peak relative to the observed X-ray band and is therefore a direct observational proxy for $\nu_\mathrm{peak}$. 
We note that an equivalent exercise can be performed in the $(\log\nu_\mathrm{peak},\,\log\nu F_\nu)$ plane and that our method applies equally to both planes.

\subsubsection{Physical motivation}

Time-dependent synchrotron emission models for blazar flares \citep{kirk1998, li2000,Perennes:2019sjx, Levy:2024eiq} predict that the hardness ratio (HR) and the total flux respond to the same underlying electron energy distribution on different timescales, because $HR$ is sensitive to the peak electron energy $\gamma_\mathrm{peak}$ while $F$ is sensitive to the total radiated power, which depends on both $\gamma_\mathrm{peak}$ and the number density of emitting electrons. The competition between particle acceleration and synchrotron cooling produces a characteristic time delay between the two variables, which is the physical origin of the hysteresis loop:

In the \emph{cooling-dominated} regime, electrons are injected rapidly and subsequently cool via synchrotron losses. Higher-energy electrons cool faster ($\dot{\gamma} \propto \gamma^2$), so $HR$ drops quickly after the flux peak while $F$ remains elevated as lower-energy electrons continue to radiate. The HR leads the flux, producing a CW loop \citep{kirk1998,tramacere2009}.

In the \emph{acceleration-dominated} regime, particle acceleration is gradual. High-energy electrons are built up slowly, so $F$ rises and peaks before the spectrum has fully hardened. The flux leads the HR, producing a CCW loop \citep{kirk1998,tramacere2009,Kapanadze:2020qpy}.

Both loop orientations are observed in the same sources at different epochs \citep{tramacere2009,fossati2000a}, and the loop direction has been used explicitly as a diagnostic to disentangle intrinsic source delays from propagation-induced effects in Lorentz invariance violation searches \citep{Perennes:2019sjx,Levy:2024eiq}.

Furthermore, the flare profiles in each variable are generally \emph{asymmetric}: the rise and decay timescales are set by different combinations of the acceleration and cooling timescales, and these combinations differ for $HR$ and $F$ because the two quantities probe different moments of the electron energy distribution \citep{fossati2000a,kirk1998}. 
Specifically, since $HR$ tracks the highest-energy electrons (which respond fastest to both acceleration and cooling), its profile tends to be more asymmetric than that of $F$. The asymmetry directions for the two variables can be opposite in the same flare: for example, in the cooling-dominated regime, $HR$ may show a fast rise and slow decay while $F$ shows the reverse. This frequency-dependent asymmetry is directly supported by observations of Markarian~421 \citep{fossati2000a} and is reproduced qualitatively by time-dependent SSC modelling \citep{kirk1998,Perennes:2019sjx}.

\subsubsection{Parametric model}

We adopt a phenomenological asymmetric Gaussian model for the flare profiles. The HR evolves as
\begin{equation}
  HR(t) = HR_0 \left[1 + A_{HR}\exp\!\left(-\frac{(t-t_{HR})^2}
  {2\sigma_{HR,\pm}^2}\right)\right],
  \label{eq:hr}
\end{equation}
and the total flux evolves as
\begin{equation}
  F(t) = F_0\left[1 + A_F\exp\!\left(-\frac{(t-t_F)^2}
  {2\sigma_{F,\pm}^2}\right)\right],
  \label{eq:flux}
\end{equation}
where $HR_0$ and $F_0$ are the quiescent values, $A_{HR}$ and
$A_F$ are the fractional flare amplitudes, $t_{HR}$ and $t_F$
are the peak times, and $\sigma_{HR,\pm}$ and $\sigma_{F,\pm}$
are the asymmetric widths:
\begin{equation}
\begin{aligned}
  \sigma_{HR,\pm} &=
  \begin{cases}
    \sigma_{HR,\mathrm{rise}} & t < t_{HR} \\
    \sigma_{HR,\mathrm{decay}} & t \geq t_{HR}
  \end{cases}\,,\\
  \sigma_{F,\pm} &=
  \begin{cases}
    \sigma_{F,\mathrm{rise}} & t < t_F \\
    \sigma_{F,\mathrm{decay}} & t \geq t_F
  \end{cases}\,.
  \label{eq:asym}
\end{aligned}
\end{equation}
The time delay between the two variables is $\Delta t = t_F - t_{HR}$. A positive $\Delta t$ (flux peaks after the HR) corresponds to the cooling-dominated regime and produces a CW loop; a negative $\Delta t$ corresponds to the acceleration-dominated regime and produces a CCW loop. In the noise-free limit, the loop area grows with $|\Delta t|$ for fixed profile shapes.

The model has seven free parameters: $A_{HR}$, $A_F$, $\sigma_{HR,\mathrm{rise}}$, $\sigma_{HR,\mathrm{decay}}$, $\sigma_{F,\mathrm{rise}}$, $\sigma_{F,\mathrm{decay}}$, and $\Delta t$. The peak time of the hardness ratio is fixed at $t_{HR} = 0$ by convention, so $\Delta t$ fully determines the relative timing of the two profiles. The number of observations $N$ is an observational choice, not a physical parameter of the model. 
The quiescent values $HR_0$ and $F_0$ set the coordinate origin and do not affect the normalised loop area $A_\mathrm{norm}$. The symmetric Gaussian model is recovered as the special case $\sigma_{HR,\mathrm{rise}} = \sigma_{HR,\mathrm{decay}} \equiv \sigma_{HR}$ and $\sigma_{F,\mathrm{rise}} = \sigma_{F,\mathrm{decay}} \equiv \sigma_F$.

Observations are taken at $N$ times within the window $[t_\mathrm{start},\, t_\mathrm{end}]$, where
\begin{align}
  t_\mathrm{start} &= \min(t_{HR}, t_F) - n_\sigma \cdot
  \max(\sigma_{HR,\mathrm{rise}},\, \sigma_{F,\mathrm{rise}}), \\
  t_\mathrm{end}   &= \max(t_{HR}, t_F) + n_\sigma \cdot
  \max(\sigma_{HR,\mathrm{decay}},\, \sigma_{F,\mathrm{decay}}),
\end{align}
and $n_\sigma$ is a fixed window multiplier (default $n_\sigma = 2$).
We investigate both uniformly and randomly spaced sampling within this window.

\subsubsection{Loop morphologies}

\begin{figure}
  \centering
  \includegraphics[width=\columnwidth]{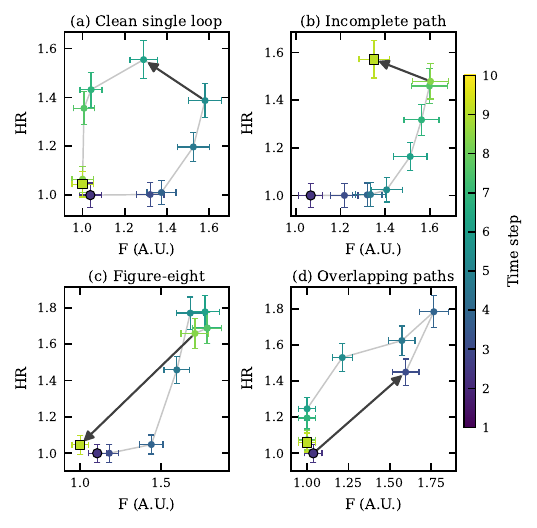}
  \caption{Representative examples of trajectory types in the $HR$--$F$ plane, generated from the asymmetric Gaussian flare model (Sect.~\ref{sec:model}) with $N = 10$ observations (circles mark the start, squares the end). From top left: (a) clean single loop with strong coherent area; (b)
incomplete trajectory where the observation window ends before the loop closes; (c) figure-eight trajectory arising from partial cancellation between two successive sub-loops; (d) nearly-degenerate trajectory with small enclosed area, where the two profiles nearly overlap. Geometric statistics for each case are given in
Table~\ref{tab:example_stats}.}
  \label{fig:signal_gallery}
\end{figure}

Fig.~\ref{fig:signal_gallery} shows representative trajectories in the $(F, HR)$ plane for different combinations of $\Delta t$ and asymmetry parameters, illustrating four typical cases: a) clean single loop, b) incomplete loop, c) figure-eight, and d) overlapping paths. 
The loop parameters are given in Table~\ref{tab:example_stats}. 

\begin{table}
  \centering
  \caption{Geometric statistics for the four example trajectories shown in Fig.~\ref{fig:signal_gallery}.}
  \label{tab:example_stats}
  \begin{tabular}{llrrrr}
    \hline\hline
    & Trajectory type
    & $A_\mathrm{norm}$ & $R_\mathrm{can}$ & $f_\mathrm{cl}$ & $d_\mathrm{cl}$ \\
    \hline
    (a) & Clean single loop & $0.97$ & $1.00$ & $0.03$ & $0.80$ \\
    (b) & Incomplete path & $0.62$ & $1.00$ & $0.38$ & $6.96$ \\
    (c) & Figure-eight & $0.40$ & $0.60$ & $0.26$ & $1.56$ \\
    (d) & Overlapping paths & $0.86$ & $1.00$ & $0.06$ & $0.97$ \\
    \hline
  \end{tabular}
\end{table}

\subsection{Null model visualization}
\label{sec:null_viz}

\begin{figure}
  \centering
  \includegraphics[width=\columnwidth]{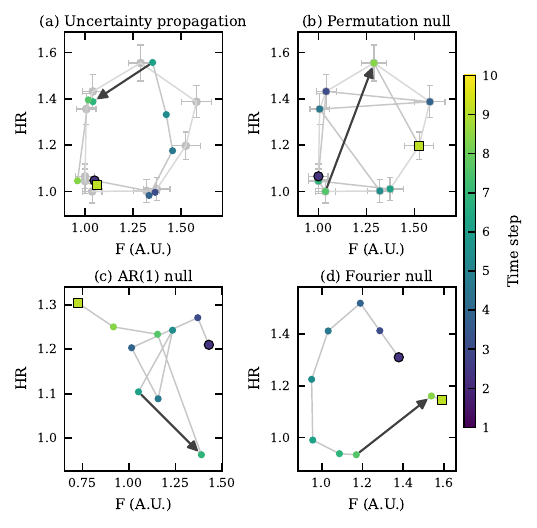}
  \caption{One uncertainty propagation of the signal (panel~a) and three null model realisations derived from the   same underlying trajectory (panels b--d). In panels~(a) and~(b), grey points with error bars show the signal for reference (panel (a) in Fig.~\ref{fig:signal_gallery}); coloured points show the noise realisation~(a), or the permuted trajectory~(b). Panels~(c) and~(d) show AR(1) and Fourier phase-randomised realisations, respectively. The geometric statistics are given in Table~\ref{tab:null_stats}.}
  \label{fig:null_gallery}
\end{figure}

Fig.~\ref{fig:null_gallery} illustrates the null model realisations derived from the case~(a) signal trajectory of Fig.~\ref{fig:signal_gallery}. Panel~(a) shows one realisation of the measurement process: the coloured points show one draw from the measurement uncertainty distribution, and the grey points with error bars show the measured signal for reference. 
Panels(b)--(d) show single realisations of the permutation, AR(1), and Fourier phase-randomised null models respectively, all derived from the noise-free signal trajectory. The grey points with error bars in panel~(b) are the original signal points, shown for direct comparison with the permuted coloured points.
The null realisations can produce trajectories that superficially resemble loops, particularly the AR(1) and Fourier nulls, which preserve the autocorrelation structure of the data, illustrating why the statistical significance test is necessary even when a loop appears visually convincing. The geometric statistics for all
four panels are given in Table~\ref{tab:null_stats}.

\begin{table}
  \centering
  \caption{Geometric statistics for the uncertainty propagation and three null model realisations shown in Fig.~\ref{fig:null_gallery}. $d_\mathrm{cl}$ is not computed for the null realisations (no uncertainties).}
  \label{tab:null_stats}
  \begin{tabular}{llrrr}
    \hline\hline
    & Trajectory type & $A_\mathrm{norm}$ & $R_\mathrm{can}$ & $f_\mathrm{cl}$ \\
    \hline
    (a) & Uncertainty propagation & $0.96$ & $0.96$ & $0.01$ \\
    (b) & Permutation null & $-0.55$ & $0.47$ & $0.14$ \\
    (c) & AR(1) null & $0.50$ & $0.96$ & $0.56$ \\
    (d) & Fourier null & $0.82$ & $0.99$ & $0.12$ \\
    \hline
  \end{tabular}
\end{table}

%%%%%%%%%%%%%%%%%%%%%%%%%%
\section{Decision framework and worked examples}
\label{sec:decision}

Sections~\ref{sec:geometry} and~\ref{sec:uncertainty} introduced geometric statistics and null models, respectively. We now combine them into a statistical analysis framework and illustrate it on the four example trajectories of Sect.~\ref{sec:model}.

\subsection{Detection criteria}
\label{sec:criteria}

A detection of statistically significant hysteresis requires all of the following to hold simultaneously.

\paragraph{Uncertainty robustness.}
When measurement uncertainties are available, the MC propagation ensemble (Sect.~\ref{sec:mcprop}) provides a distribution of $A_\mathrm{norm}$ values consistent with the data. A detection is robust against measurement noise if this distribution is concentrated on one side of zero. If the distribution has significant support on both sides of zero, the measured $|A_\mathrm{norm}|$ is consistent with zero and hysteresis has no support.  

\paragraph{Statistical significance.} 
As discussed in Sect.~\ref{sec:null_motivation}, statistically significant detection of $|A_\mathrm{norm}|$ does not necessarily imply an underlying physical process that causes hysteresis-like behaviour. 
Therefore, the method computes four $p$-values: $p_\mathrm{perm}$, $p_\mathrm{AR}$, $p_\mathrm{Fourier}$, and $p_\mathrm{full}$ from the combined null ensemble. 
We leave it to the analysers to decide the appropriate threshold based on the requirements of their analysis. Their physical interpretation follows the hierarchy established in Sect.~\ref{sec:combined}: a small $p_\mathrm{perm}$ indicates the loop is unlikely to arise from a different temporal ordering of the same data; a small $p_\mathrm{AR}$ indicates it is unlikely to arise from correlated red-noise variability; a small $p_\mathrm{Fourier}$ indicates it is unlikely to arise from a stochastic process with the same power spectrum. Only when all three individual $p$-values are small can one argue that the loop reflects a genuine physical coupling between the two observables, rather than an artefact of the autocorrelation structure or sampling.

\subsection{Diagnostics}
\label{sec:diagnostics}

In addition to the detection criteria, we identify several diagnostic tools. 

\paragraph{Sample size.}
We require $N \geq 4$. Three points define a single triangle with a trivially non-zero signed area. With $N = 4$, two triangles can have opposite signs, providing the minimum number of points for a non-degenerate hysteresis measurement. The reliability of the null distributions and the detection power both improve with increasing $N$. We do not impose a stricter minimum, but note that the number of distinct permutations is $N! \leq 720$ for $N \leq 6$, which limits the statistical resolution of $p_\mathrm{perm}$ at small sample sizes. 
In addition, we remind the reader that the Fourier null requires at least $N = 6$ observations (\ref{sec:fouriersurr}).

\paragraph{Cancellation diagnostics.}
$R_\mathrm{cancel}$ is reported as a diagnostic alongside the $p$-values. Values significantly below 1 indicate partial cancellation between sub-loops of opposite orientation, as in the figure-eight case (Fig.~\ref{fig:signal_gallery}c). In such cases, the signed area $A_\mathrm{norm}$ does not represent a single coherent loop. 
In such cases, one might consider restricting the analysis to a subset of the observed data. 
In either case, the physical interpretation requires care, even if the $p$-values are small.

\begin{figure}
  \centering
  \includegraphics[width=\columnwidth]{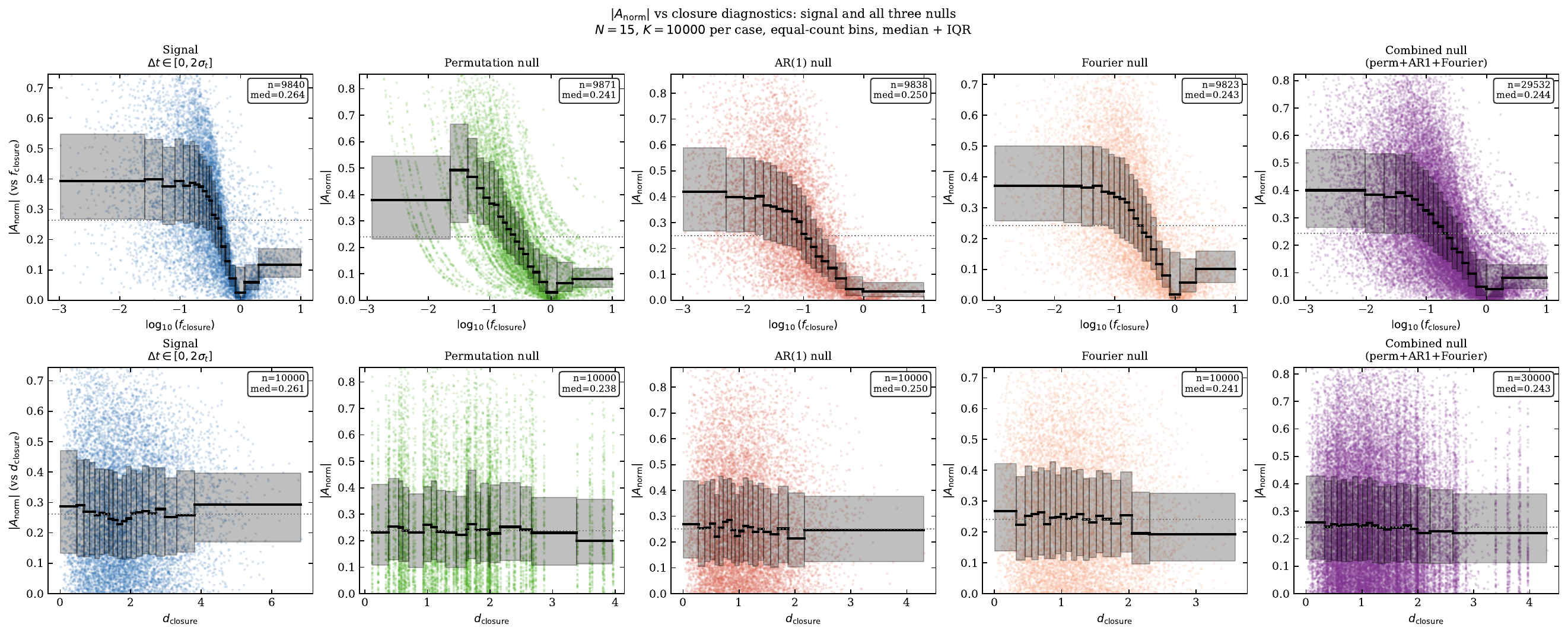}
  \caption{Median $|A_\mathrm{norm}|$ and interquartile range per equal-count bin as a function of $\log_{10}(f_\mathrm{cl})$ (top row) and $d_\mathrm{cl}$ (bottom row) for different null trajectories at $N = 15$ ($K = 10{,}000$ realisations). The dotted horizontal line marks the overall median $|A_\mathrm{norm}|$.}
  \label{fig:closure}
\end{figure}

\paragraph{Closure diagnostics.}
Neither $f_\mathrm{cl}$ nor $d_\mathrm{cl}$ is used as a hard detection criterion. Simulations show that $f_\mathrm{cl}$ has a very wide scatter at any given value of $|A_\mathrm{norm}|$, with limited predictive power (Fig.~\ref{fig:closure}), while $d_\mathrm{cl}$ shows essentially no trend with $|A_\mathrm{norm}|$ in the null distribution. Both are reported as diagnostics alongside the primary statistics. Large values indicate that the considered trajectory did not cover a complete cycle, whether because the observations did not extend over the whole cycle of the governing physical process or because the process is such that the observed parameters do not form a closed loop. 

\paragraph{}
Although none of these diagnostic tools are used as detection criteria, they should be considered when forming a physical interpretation of the results, particularly when more than one effect is visible in the data sample and especially so when only a subset of data are being analysed.

\subsection{Worked examples}
\label{sec:worked}

\begin{figure}
  \centering
  \includegraphics[width=\columnwidth]{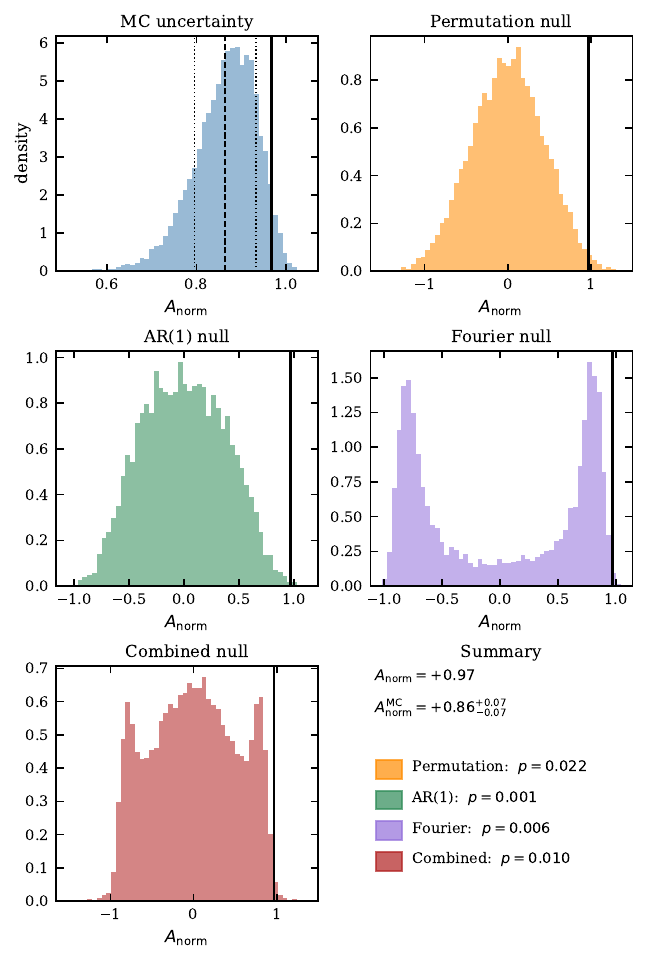}
  \caption{Hysteresis analysis diagnostics for the clean single-loop trajectory (case~a). Each panel shows the distribution of $A_\mathrm{norm}$ values from a different null ensemble ($K = 10^4$ realisations each). The solid vertical line marks the observed $A_\mathrm{norm}$. \textit{Top left:} MC uncertainty propagation; the dashed line shows the MC mean and the dotted lines the $1\sigma$ CI.\textit{Top right, middle left, middle right:} permutation, AR(1), and Fourier phase-randomisation null distributions. \textit{Bottom left:} combined null (concatenation of all three ensembles). \textit{Bottom right:} summary of the observed $A_\mathrm{norm}$ with its MC uncertainty and empirical two-sided $p$-values against each null model; coloured bars correspond to the histograms in the respective panels.}
  \label{fig:worked}
\end{figure}

Table~\ref{tab:worked} shows the results of applying the full framework to the four signal trajectories of Fig.~\ref{fig:signal_gallery}. For each trajectory, we report the geometric statistics, the MC CI of $A_\mathrm{norm}$ from $K_\mathrm{MC} = 10^4$ uncertainty propagation realisations, and all four $p$-values computed from $K = 10^4$ null model realisations.
The position of $|A_\mathrm{norm}|$ with respect to the statistical distributions for case~(a) is given in Fig.~\ref{fig:worked}. 

\begin{table*}
\centering
\caption{Hysteresis analysis results. Columns: trajectory label; normalised signed area $A_\mathrm{norm}$; 1-sigma MC CI on $A_\mathrm{norm}$; loop orientation (CCW/CW); cancellation ratio $R_\mathrm{can}$; closure fraction $f_\mathrm{cl}$; endpoint separation $d_\mathrm{cl}$; empirical two-sided $p$-values against the permutation; AR(1); Fourier; and combined null ensembles.}
\label{tab:worked}
\begin{tabular}{lcccccccccc}
\hline\hline
Case & $A_\mathrm{norm}$ & MC $1\sigma$ & Orient. & $R_\mathrm{can}$ & $f_\mathrm{cl}$ & $d_\mathrm{cl}$ & $p_\mathrm{perm}$ & $p_\mathrm{AR}$ & $p_\mathrm{Fourier}$ & $p_\mathrm{full}$ \\
\hline
(a) & 0.97 & [0.80, 0.93] & CCW & 1.00 & 0.03 & 0.80 & 0.022 & 0.001 & 0.006 & 0.010 \\
(b) & 0.62 & [0.43, 0.60] & CCW & 1.00 & 0.38 & 6.96 & 0.121 & 0.075 & 0.365 & 0.187 \\
(c) & 0.40 & [0.10, 0.55] & CCW & 0.60 & 0.26 & 1.56 & 0.352 & 0.324 & 0.767 & 0.481 \\
(d) & 0.86 & [0.60, 0.88] & CCW & 1.00 & 0.06 & 0.97 & 0.022 & 0.010 & 0.096 & 0.043 \\
\hline
\end{tabular}
\end{table*}

Case~(a)---the clean single loop---yields $A_\mathrm{norm} = +0.97$, with a narrow MC interval $[+0.80, +0.93]$ lying entirely on the positive (CCW) side. The small $p$-values under all three null models ($p_\mathrm{perm} = 0.022$, $p_\mathrm{AR} = 0.001$, $p_\mathrm{Fourier} = 0.006$, $p_\mathrm{full} = 0.010$) confirm that the observed loop area cannot plausibly arise from random reordering, from correlated red-noise variability, or from a stochastic process with the same power spectrum. The small $f_\mathrm{cl} = 0.03$ and $d_\mathrm{cl} = 0.80$ confirm that the trajectory covers a complete cycle. This is a clear detection of hysteresis. 

Case~(b)---the incomplete path---has $A_\mathrm{norm} = +0.62$, with the MC interval $[+0.43, +0.60]$ lying below the observed value, but still significantly distant from zero, confirming that the loop area is robust to measurement noise. However, all $p$-values are large ($p_\mathrm{perm} = 0.121$, $p_\mathrm{AR} = 0.075$, $p_\mathrm{Fourier} = 0.365$, $p_\mathrm{full} = 0.187$), meaning the observed $A_\mathrm{norm}$ cannot be clearly distinguished from what the null models can produce. The large $f_\mathrm{cl} = 0.38$ and $d_\mathrm{cl} = 6.96$ indicate an open path. Whether this reflects an incomplete observation window or a physical process that genuinely does not close the loop cannot be determined from the data alone, and requires physical knowledge of the system. Note that an open-path trajectory is not necessarily less physically meaningful than a closed one, as it may still reflect a genuine coupling between the two observables. However, the current sample does not provide sufficient statistical evidence to distinguish it from the null hypotheses. 

Case~(c)---the figure-eight---yields $A_\mathrm{norm} = +0.40$ with $R_\mathrm{cancel} = 0.60$, indicating significant cancellation between two sub-loops of opposite orientation. The wide MC interval $[+0.10, +0.55]$ reflects the sensitivity of the signed area to noise in a self-crossing trajectory. All $p$-values are large ($p_\mathrm{full} = 0.481$), consistent with no net hysteresis signal. Even if the individual sub-loops were significant, their contributions to the total signed area $A_\mathrm{norm}$ would cancel, preventing detection of a single coherent hysteresis pattern. In such cases, the data should be examined for the presence of multiple overlapping physical processes before proceeding with the detection claim. 

Case~(d)---a trajectory in which the two arms lie close together, which may represent either weak hysteresis or a statistical fluctuation that mimics hysteresis-like behaviour---yields $A_\mathrm{norm} = +0.86$, which on its own would suggest a large loop area. However, the MC interval $[+0.60, +0.88]$ is noticeably wider than for cases~(a) and~(b), reflecting the greater sensitivity of the enclosed area to measurement noise when the two arms of the trajectory are not well separated: small perturbations can easily shift points across the narrow gap, significantly changing the apparent area.

%%%%%%%%%%%%%%%%%%%%%%%%%%
\section{Application to real data: Markarian~421}
\label{sec:application}

To demonstrate the applicability of the framework to real data, we apply it to an XMM-Newton observation of the high-synchrotron-peaked BL Lac object Markarian~421 ($z = 0.031$) obtained on 2023 December 13 during a period of enhanced X-ray and very-high-energy (VHE) gamma-ray activity \citep{MAGIC:2024mjz}. Markarian~421 is the archetypical TeV blazar and one of the most intensively monitored sources for spectral hysteresis studies \citep{fossati2000a, tramacere2009, Levy:2024eiq}.

\citet{MAGIC:2024mjz} present a multi-wavelength campaign covering radio to VHE gamma rays, including simultaneous X-ray polarisation measurements with IXPE. During the XMM-Newton observation, a CCW hysteresis loop was identified in the plane of X-ray HR versus 2--10\,keV flux (Fig.~6 in~\citealt{MAGIC:2024mjz}), which was interpreted as evidence that the X-ray emission is dominated by particles near the high-energy cutoff of the electron distribution. We apply our framework to quantify the statistical significance of this loop.

\subsection{Data}
\label{sec:app_data}

\begin{figure}
  \centering
  \includegraphics[width=\columnwidth]{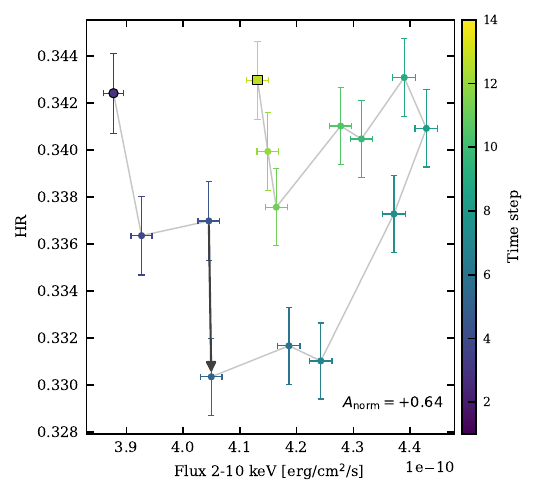}
  \caption{Hysteresis analysis of Markarian~421 from the XMM-Newton observation on 2023 December 13 \citep{MAGIC:2024mjz}. The trajectory in the HR--flux plane is shown, with colour encoding the time step (1 to 14), and the square symbol marking the starting point. The CCW loop is clearly visible. The observed $A_\mathrm{norm} = +0.64$ and
  the complete statistical results are given in Table~\ref{tab:app_results} and Fig.~\ref{fig:mrk421_diag}.}
  \label{fig:mrk421}
\end{figure}

The XMM-Newton light curve from 2023 December 13 was provided in binned form by \citet{MAGIC:2024mjz}. The dataset consists of $N = 14$ time bins covering the full observation. For each bin we use the 2--10\,keV flux as the independent variable and the HR $\mathrm{HR} = F_{2\text{--}10\,\mathrm{keV}} / F_{0.3\text{--}2\,\mathrm{keV}}$ as the dependent variable. Measurement uncertainties in both coordinates are propagated from the individual band flux uncertainties. The flux ranges from $3.88 \times 10^{-10}$ to $4.43 \times 10^{-10}$ erg cm$^{-2}$ s$^{-1}$, and the HR from 0.330 to 0.343. The loop is shown in Fig.~\ref{fig:mrk421}.

\subsection{Results}
\label{sec:app_results}

\begin{figure}
  \centering
  \includegraphics[width=\columnwidth]{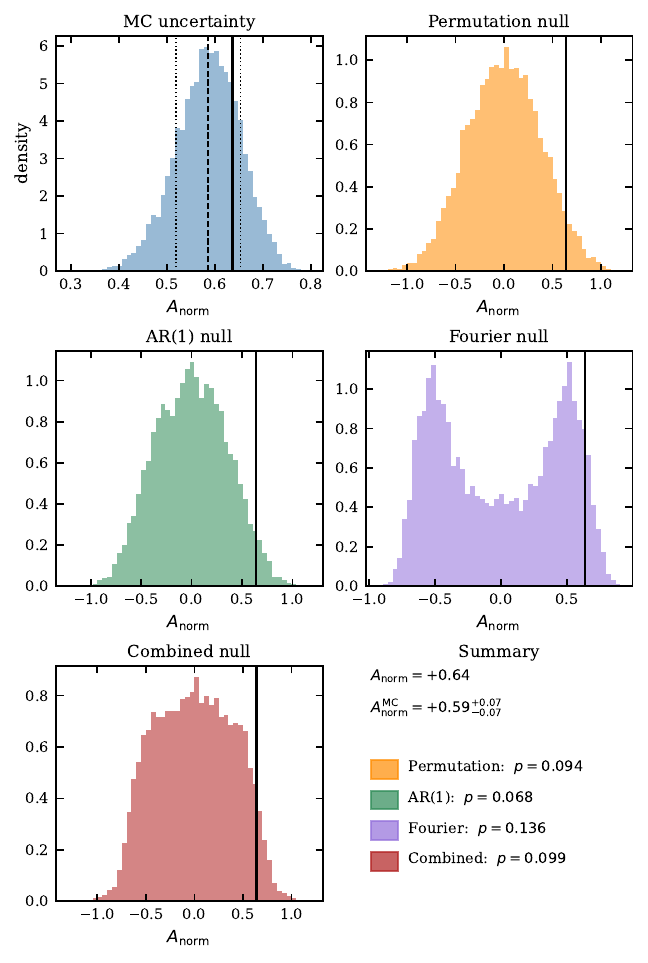}
  \caption{Statistical diagnostics for the Markarian~421 XMM-Newton trajectory. Each panel shows the distribution of $A_\mathrm{norm}$ values from a different null ensemble ($K = 10^4$ realisations each). The solid vertical line marks the observed $A_\mathrm{norm} = +0.64$. \textit{Top left:} MC uncertainty propagation; the dashed line shows the MC mean and the dotted lines the $1\sigma$ interval. \textit{Top right, middle left, middle right:} permutation, AR(1), and Fourier null distributions. \textit{Bottom left:} combined null. \textit{Bottom right:} summary of the observed $A_\mathrm{norm}$ with its MC uncertainty and empirical two-sided $p$-values.}
  \label{fig:mrk421_diag}
\end{figure}

We apply \texttt{analyse\_hysteresis.py} to this dataset using $K_\mathrm{null} = 10^4$ realisations per null model and $K_\mathrm{MC} = 10^4$ MC uncertainty propagation realisations. The results are summarised in Table~\ref{tab:app_results} and Fig.~\ref{fig:mrk421_diag}.

\begin{table}
  \caption{Hysteresis analysis results for the XMM-Newton observation of Markarian~421 on 2023 December 13, from \citet{MAGIC:2024mjz}.}
  \label{tab:app_results}
  \centering
  \begin{tabular}{ll}
    \hline\hline
    Statistic & Value \\
    \hline
    $N$                              & 14 \\
    $A_\mathrm{norm}$                & $+0.64$ \\
    MC $1\sigma$ interval            & $[+0.52,\, +0.65]$ \\
    Orientation                      & CCW \\
    $R_\mathrm{can}$                 & 0.92 \\
    $f_\mathrm{cl}$                  & 0.17 \\
    $d_\mathrm{cl}$                  & 9.68 \\
    $p_\mathrm{perm}$                & 0.094 \\
    $p_\mathrm{AR}$                  & 0.068 \\
    $p_\mathrm{Fourier}$             & 0.136 \\
    $p_\mathrm{full}$                & 0.099 \\
    \hline
  \end{tabular}
  \tablefoot{$K = 10^4$ realisations per null model; $K_\mathrm{MC} = 10^4$ uncertainty propagation realisations.}
\end{table}

The analysis yields $A_\mathrm{norm} = +0.64$, confirming a CCW loop in the HR--flux plane, consistent with the visual identification in \citet{MAGIC:2024mjz}. The MC uncertainty interval $[+0.52, +0.65]$ is entirely positive and well separated from zero, confirming that the CCW orientation and the non-zero loop area are robust against measurement noise. The cancellation ratio $R_\mathrm{can} = 0.92$ indicates a single coherent loop with negligible sub-loop cancellation.

The trajectory is open. $f_\mathrm{cl} = 0.17$ and $d_\mathrm{cl} = 9.68$ indicate that the observation does not cover a complete cycle, or the process does not close a full loop. 
All three null model $p$-values are in the range 0.068--0.136, and $p_\mathrm{full} = 0.099$. These values are sub-threshold at conventional significance levels (e.g.\ $\alpha = 0.05$). However, as discussed in Sect.~\ref{sec:opentrajectories}, the open trajectory geometry reduces the detection power, and the result should be interpreted in this context. 
AR(1) gives the smallest individual $p$-value ($p_\mathrm{AR} = 0.068$), meaning the observed loop is difficult to reproduce with autocorrelated red-noise surrogates. 
The Fourier $p$-value ($p_\mathrm{Fourier} = 0.136$) is the largest of the three. At $N = 14$ the power spectrum estimate carries non-negligible sampling uncertainty, and $p_\mathrm{Fourier}$ should be interpreted with corresponding caution (Sect.~\ref{sec:fouriersurr}).

The CCW orientation is physically interpretable within the framework of \citet{tramacere2009}: it corresponds to the acceleration timescale being shorter than the cooling timescale, so the spectral hardening precedes the flux rise. 
\citet{MAGIC:2024mjz} attribute the CCW loop to X-ray emission from electrons near the high-energy cutoff of the distribution, consistent with an extreme configuration of the Turbulent Extreme Multi-Zone model. Our quantitative analysis supports this interpretation while clarifying the statistical evidence: the loop is robustly detected against measurement noise, but the formal $p$-values against stochastic null models remain above $\alpha = 0.05$, primarily because of the limited sample size and open trajectory geometry.

%%%%%%%%%%%%%%%%%%%%%%%%%%
\section{Conclusions}
\label{sec:conclusions}

We have presented a statistically robust framework for detecting, quantifying, and classifying hysteresis patterns in ordered two-dimensional astrophysical data with measurement uncertainties. The framework is entirely model-independent and applicable to any pair of simultaneously measured observables.

The primary detection statistic is the normalised signed area $A_\mathrm{norm}$, defined as the open-path shoelace area of the chronologically ordered trajectory divided by the area of its convex hull. This normalisation makes the statistic translation- and rotation-invariant and enables scale-free comparison between datasets. The decomposition of the total area into an open-path contribution $A_\mathrm{open}$ and an artificial closure contribution $A_\mathrm{closure}$ separates the physically meaningful hysteresis signal from the geometrical artefact introduced by forcing trajectory closure. The complementary diagnostics $R_\mathrm{can}$, $f_\mathrm{cl}$, and $d_\mathrm{cl}$ characterise, respectively, the coherence of the loop (whether the trajectory is a single loop or a self-cancelling multi-loop structure), the fraction of the enclosed area attributable to forced closure, and the statistical significance of the endpoint separation relative to measurement uncertainties. We verified through simulation that neither $f_\mathrm{cl}$ nor $d_\mathrm{cl}$ have sufficient predictive power for $|A_\mathrm{norm}|$ to serve as a hard detection criterion. However, they are both reported as diagnostics only.

Measurement uncertainties are propagated into $A_\mathrm{norm}$ via MC sampling, yielding a distribution whose mean and 1$\sigma$ CI characterise the robustness of the observed loop to noise. Statistical significance is assessed against three null models that represent qualitatively different stochastic mechanisms: random permutation of the time ordering (testing whether the loop could arise from any arrangement of the same data points), AR(1) surrogates (testing against correlated red-noise variability), and Fourier phase-randomised surrogates (testing against any process with the same power spectrum as the observed data). The three individual $p$-values carry distinct physical meaning, and we recommend reporting all three alongside the combined $p_\mathrm{full}$ from the pooled null ensemble. A small $p_\mathrm{full}$ indicates that the observed loop is difficult to explain under any of the three null hypotheses simultaneously, while the individual $p$-values diagnose which specific stochastic mechanisms can be excluded.

We note that the Fourier null distribution is bimodal for some trajectories, depending on their spectral content, and that all three null distributions can rarely produce $|A_\mathrm{norm}| > 1$ due to self-intersecting (double-winding) surrogates. Both effects are quantified and shown to have negligible impact on the reported $p$-values at sample sizes typical of the present application.

The framework was validated on synthetic blazar flare trajectories in the $HR$--$F$ plane, generated from an asymmetric Gaussian flare model physically motivated by the different timescales of particle acceleration and synchrotron cooling in HBL sources. Four representative trajectory types---a clean single loop, an incomplete path, a figure-eight, and a near-degenerate trajectory---were analysed, illustrating how the combination of $A_\mathrm{norm}$, its MC uncertainty interval, and the three $p$-values jointly characterise statistical evidence and physical interpretation.

As a demonstration on real data, we applied the framework to an XMM-Newton observation of Markarian~421 during a December 2023 flaring episode studied by the MAGIC Collaboration \citep{MAGIC:2024mjz}. The analysis yields $A_\mathrm{norm} = +0.64$ with the MC interval $[+0.52, +0.65]$ entirely positive, confirming a CCW loop that is robust against measurement noise. 
The $p$-values against all three null models fall in the range 0.068--0.136, with $p_\mathrm{full} = 0.099$. The open trajectory geometry ($d_\mathrm{cl} = 9.68$) reduces the detection power. Additionally, the Fourier null distribution is bimodal for this trajectory, which can affect the interpretation of $p_\mathrm{Fourier}$ (Sect.~\ref{sec:fouriersurr}). 
The CCW orientation is physically consistent with the acceleration-dominated particle injection regime described by \citet{tramacere2009}, in agreement with the interpretation of \citet{MAGIC:2024mjz}. 

Fully documented open-source Python software that implements the complete framework is available at \url{https://github.com/tterzic/hysteresis-detection}.

%%%%%%%%%%%%%%%%%%%%%%%%%%
\section*{Data availability}
The materials necessary to reproduce the figures and other results in this work are available in the project GitHub repository \url{https://github.com/tterzic/hysteresis-detection}. 

%%%%%%%%%%%%%%%%%%%%%%%%%%
\begin{acknowledgements}
The author would like to acknowledge the support by the University of Rijeka through the project uniri-iz-25-119 and the Croatian Science Foundation (HrZZ) Project IP-2022-10-4595, and the networking support by CA23130 BridgeQG (Bridging high and low energies in search of Quantum Gravity, \url{https://www.cost.eu/actions/CA23130/}).    

During the method development and paper drafting, it was brought to our attention that a similar approach was developed by Guillaume Grolleron. The two studies were conducted simultaneously but independently. 
\end{acknowledgements}

%%%%%%%%%%%%%%%%%%%%%%%%%%
\bibliographystyle{aa}
\bibliography{MeasuringHysteresis}

@article{1996ApJ...470L..89T,
    author = "Takahashi, T. and Tashiro, M. and Madejski, G. and Kubo, H. and Kamae, T. and Kataoka, J. and Kii, T. and Makino, F. and Makishima, K. and Yamasaki, N.",
    title = "{ASCA Observation of an X-Ray/TeV Flare from the BL Lacertae Object Markarian 421}",
    doi = "10.1086/310302",
    journal = "Astrophys. J. Lett.",
    volume = "470",
    pages = "L89",
    year = "1996"
}

@article{2000ApJ...541..166F,
    author = "Fossati, G. and Celotti, A. and Chiaberge, M. and Zhang, Y. H. and Chiappetti, L. and Ghisellini, G. and Maraschi, L. and Tavecchio, F. and Pian, E. and Treves, A.",
    title = "{X-ray emission of mkn 421: new clues from its spectral evolution. 2. spectral analysis and physical constraints}",
    eprint = "astro-ph/0005067",
    archivePrefix = "arXiv",
    doi = "10.1086/309430",
    journal = "Astrophys. J.",
    volume = "541",
    pages = "166",
    year = "2000"
}

@ARTICLE{2012SoPh..276..407S,
       author = {{Suyal}, Vinita and {Prasad}, Awadhesh and {Singh}, Harinder P.},
        title = "{Hysteresis in a Solar Activity Cycle}",
      journal = {\solphys},
         year = 2012,
        month = feb,
       volume = {276},
       number = {1-2},
        pages = {407-414},
          doi = {10.1007/s11207-011-9889-0},
archivePrefix = {arXiv},
       eprint = {1112.5236},
 primaryClass = {astro-ph.SR},
       adsurl = {https://ui.adsabs.harvard.edu/abs/2012SoPh..276..407S}
}

@ARTICLE{2003NewA....8..745O,
       author = {{{\"O}zg{\"u}{\c{c}}}, A. and {Ata{\c{c}}}, T.},
        title = "{Effects of hysteresis in solar cycle variations between flare index and cosmic rays}",
      journal = {\na},
         year = 2003,
        month = nov,
       volume = {8},
       number = {8},
        pages = {745-750},
          doi = {10.1016/S1384-1076(03)00063-0},
       adsurl = {https://ui.adsabs.harvard.edu/abs/2003NewA....8..745O}
}

@ARTICLE{1994SoPh..150..347B,
       author = {{Bachmann}, Kurt T. and {White}, Oran R.},
        title = "{Observations of Hysteresis in Solar Cycle Variations among Seven Solar Activity Indicators}",
      journal = {\solphys},
         year = 1994,
        month = mar,
       volume = {150},
       number = {1-2},
        pages = {347-357},
          doi = {10.1007/BF00712896},
       adsurl = {https://ui.adsabs.harvard.edu/abs/1994SoPh..150..347B}
}

@ARTICLE{2017ApJ...834....2A,
       author = {{Abeysekara}, A.~U. and {Archambault}, S. and {Archer}, A. and {Benbow}, W. and {Bird}, R. and {Buchovecky}, M. and {Buckley}, J.~H. and {Bugaev}, V. and {Cardenzana}, J.~V. and {Cerruti}, M. and {Chen}, X. and {Ciupik}, L. and {Connolly}, M.~P. and {Cui}, W. and {Eisch}, J.~D. and {Falcone}, A. and {Feng}, Q. and {Finley}, J.~P. and {Fleischhack}, H. and {Flinders}, A. and {Fortson}, L. and {Furniss}, A. and {Griffin}, S. and {H{\r{a}}kansson}, N. and {Hanna}, D. and {Hervet}, O. and {Holder}, J. and {Humensky}, T.~B. and {H{\"u}tten}, M. and {Kaaret}, P. and {Kar}, P. and {Kertzman}, M. and {Kieda}, D. and {Krause}, M. and {Kumar}, S. and {Lang}, M.~J. and {Maier}, G. and {McArthur}, S. and {McCann}, A. and {Meagher}, K. and {Moriarty}, P. and {Mukherjee}, R. and {Nieto}, D. and {O'Brien}, S. and {Ong}, R.~A. and {Otte}, A.~N. and {Park}, N. and {Pelassa}, V. and {Pohl}, M. and {Popkow}, A. and {Pueschel}, E. and {Ragan}, K. and {Reynolds}, P.~T. and {Richards}, G.~T. and {Roache}, E. and {Sadeh}, I. and {Santander}, M. and {Sembroski}, G.~H. and {Shahinyan}, K. and {Staszak}, D. and {Telezhinsky}, I. and {Tucci}, J.~V. and {Tyler}, J. and {Wakely}, S.~P. and {Weinstein}, A. and {Wilhelm}, A. and {Williams}, D.~A. and {VERITAS Collaboration} and {Ahnen}, M.~L. and {Ansoldi}, S. and {Antonelli}, L.~A. and {Antoranz}, P. and {Arcaro}, C. and {Babic}, A. and {Banerjee}, B. and {Bangale}, P. and {Barres de Almeida}, U. and {Barrio}, J.~A. and {Becerra Gonz{\'a}lez}, J. and {Bednarek}, W. and {Bernardini}, E. and {Berti}, A. and {Biasuzzi}, B. and {Biland}, A. and {Blanch}, O. and {Bonnefoy}, S. and {Bonnoli}, G. and {Borracci}, F. and {Bretz}, T. and {Carosi}, R. and {Carosi}, A. and {Chatterjee}, A. and {Colin}, P. and {Colombo}, E. and {Contreras}, J.~L. and {Cortina}, J. and {Covino}, S. and {Cumani}, P. and {Da Vela}, P. and {Dazzi}, F. and {De Angelis}, A. and {De Lotto}, B. and {de O{\~n}a Wilhelmi}, E. and {Di Pierro}, F. and {Doert}, M. and {Dom{\'\i}nguez}, A. and {Dominis Prester}, D. and {Dorner}, D. and {Doro}, M. and {Einecke}, S. and {Eisenacher Glawion}, D. and {Elsaesser}, D. and {Engelkemeier}, M. and {Fallah Ramazani}, V. and {Fern{\'a}ndez-Barral}, A. and {Fidalgo}, D. and {Fonseca}, M.~V. and {Font}, L. and {Fruck}, C. and {Galindo}, D. and {Garc{\'\i}a L{\'o}pez}, R.~J. and {Garczarczyk}, M. and {Gaug}, M. and {Giammaria}, P. and {Godinovi{\'c}}, N. and {Gora}, D. and {Guberman}, D. and {Hadasch}, D. and {Hahn}, A. and {Hassan}, T. and {Hayashida}, M. and {Herrera}, J. and {Hose}, J. and {Hrupec}, D. and {Hughes}, G. and {Idec}, W. and {Kodani}, K. and {Konno}, Y. and {Kubo}, H. and {Kushida}, J. and {Lelas}, D. and {Lindfors}, E. and {Lombardi}, S. and {Longo}, F. and {L{\'o}pez}, M. and {L{\'o}pez-Coto}, R. and {Majumdar}, P. and {Makariev}, M. and {Mallot}, K. and {Maneva}, G. and {Manganaro}, M. and {Mannheim}, K. and {Maraschi}, L. and {Marcote}, B. and {Mariotti}, M. and {Mart{\'\i}nez}, M. and {Mazin}, D. and {Menzel}, U. and {Mirzoyan}, R. and {Moralejo}, A. and {Moretti}, E. and {Nakajima}, D. and {Neustroev}, V. and {Niedzwiecki}, A. and {Nievas Rosillo}, M. and {Nilsson}, K. and {Nishijima}, K. and {Noda}, K. and {Nogu{\'e}s}, L. and {N{\"o}the}, M. and {Paiano}, S. and {Palacio}, J. and {Palatiello}, M. and {Paneque}, D. and {Paoletti}, R. and {Paredes}, J.~M. and {Paredes-Fortuny}, X. and {Pedaletti}, G. and {Peresano}, M. and {Perri}, L. and {Persic}, M. and {Poutanen}, J. and {Prada Moroni}, P.~G. and {Prandini}, E. and {Puljak}, I. and {Garcia}, J.~R. and {Reichardt}, I. and {Rhode}, W. and {Rib{\'o}}, M. and {Rico}, J. and {Saito}, T. and {Satalecka}, K. and {Schroeder}, S. and {Schweizer}, T. and {Shore}, S.~N. and {Sillanp{\"a}{\"a}}, A. and {Sitarek}, J. and {Snidaric}, I. and {Sobczynska}, D. and {Stamerra}, A.},
        title = "{A Search for Spectral Hysteresis and Energy-dependent Time Lags from X-Ray and TeV Gamma-Ray Observations of Mrk 421}",
      journal = {\apj},
         year = 2017,
        month = jan,
       volume = {834},
       number = {1},
          eid = {2},
        pages = {2},
          doi = {10.3847/1538-4357/834/1/2},
archivePrefix = {arXiv},
       eprint = {1611.04626},
 primaryClass = {astro-ph.HE},
       adsurl = {https://ui.adsabs.harvard.edu/abs/2017ApJ...834....2A}
}

@ARTICLE{miyamoto1995,
  author  = {Miyamoto, S. and Kitamoto, S. and Hayashida, K. and Egoshi, W.},
  title   = {Large hysteretic behavior of stellar black hole candidate
             X-ray binaries},
  journal = {\apjl},
  year    = {1995},
  volume  = {442},
  pages   = {L13--L16},
  doi     = {10.1086/187804}
}

@ARTICLE{zdziarski2004,
  author  = {Zdziarski, A.~A. and Gierli{\'n}ski, M. and
             Miko{\l}ajewska, J. and others},
  title   = {Statistical properties of the variability of
             {XTE~J1550$-$564}},
  journal = {\mnras},
  year    = {2004},
  volume  = {351},
  pages   = {791--807},
  doi     = {10.1111/j.1365-2966.2004.07830.x}
}

@ARTICLE{belloni2005,
  author  = {Belloni, T. and Homan, J. and Casella, P. and others},
  title   = {The evolution of the timing properties of the black-hole
             transient {GX~339-4} during its 2002/2003 outburst},
  journal = {\aap},
  year    = {2005},
  volume  = {440},
  pages   = {207--222},
  doi     = {10.1051/0004-6361:20042457}
}

@ARTICLE{uttley2002,
  author  = {Uttley, P. and McHardy, I.~M. and Papadakis, I.~E.},
  title   = {Measuring the broad-band power spectra of active
             galactic nuclei with {RXTE}},
  journal = {\mnras},
  year    = {2002},
  volume  = {332},
  pages   = {231--248},
  doi     = {10.1046/j.1365-8711.2002.05298.x}
}

@ARTICLE{vaughan2003a,
  author  = {Vaughan, S. and Edelson, R. and Warwick, R.~S. and
             Uttley, P.},
  title   = {On characterizing the variability properties of
             {X}-ray light curves from active galaxies},
  journal = {\mnras},
  year    = {2003},
  volume  = {345},
  pages   = {1271--1284},
  doi     = {10.1046/j.1365-8711.2003.07042.x}
}

@ARTICLE{mchardy2004,
  author  = {McHardy, I.~M. and Papadakis, I.~E. and Uttley, P. and
             Page, M.~J. and Mason, K.~O.},
  title   = {Combined long and short time-scale {X}-ray variability
             of {NGC~4051}},
  journal = {\mnras},
  year    = {2004},
  volume  = {348},
  pages   = {783--801},
  doi     = {10.1111/j.1365-2966.2004.07376.x}
}

@ARTICLE{uttley2005,
  author  = {Uttley, P. and McHardy, I.~M. and Vaughan, S.},
  title   = {Non-linear {X}-ray variability in {X}-ray binaries
             and active galaxies},
  journal = {\mnras},
  year    = {2005},
  volume  = {359},
  pages   = {345--362},
  doi     = {10.1111/j.1365-2966.2005.08886.x}
}

@ARTICLE{timmer1995,
  author  = {Timmer, J. and K{\"o}nig, M.},
  title   = {On generating power law noise},
  journal = {\aap},
  year    = {1995},
  volume  = {300},
  pages   = {707--710}
}

@ARTICLE{emmanoulopoulos2013,
  author  = {Emmanoulopoulos, D. and McHardy, I.~M. and
             Papadakis, I.~E.},
  title   = {Generating artificial light curves: revisited and
             updated},
  journal = {\mnras},
  year    = {2013},
  volume  = {433},
  pages   = {907--927},
  doi     = {10.1093/mnras/stt764}
}

@ARTICLE{corbel2013,
  author  = {Corbel, S. and Coriat, M. and Brocksopp, C. and others},
  title   = {The `universal' radio/X-ray flux correlation: the case
             study of the black hole {GX~339-4}},
  journal = {\mnras},
  year    = {2013},
  volume  = {428},
  pages   = {2500--2515},
  doi     = {10.1093/mnras/sts215}
}

@ARTICLE{dunn2010,
  author  = {Dunn, R.~J.~H. and Fender, R.~P. and K{\"o}rding, E.~G.
             and Belloni, T. and Cabanac, C.},
  title   = {A global study of the behaviour of black hole X-ray
             binary discs},
  journal = {\mnras},
  year    = {2010},
  volume  = {403},
  pages   = {61--82},
  doi     = {10.1111/j.1365-2966.2010.16114.x}
}

@ARTICLE{akaike1974,
  author  = {Akaike, H.},
  title   = {A new look at the statistical model identification},
  journal = {IEEE Trans. Autom. Control},
  year    = {1974},
  volume  = {19},
  pages   = {716--723},
  doi     = {10.1109/TAC.1974.1100705}
}

@BOOK{davison1997,
  author    = {Davison, A.~C. and Hinkley, D.~V.},
  title     = {Bootstrap Methods and their Application},
  publisher = {Cambridge University Press},
  year      = {1997}
}

@BOOK{brockwell2002,
  author    = {Brockwell, P.~J. and Davis, R.~A.},
  title     = {Introduction to Time Series and Forecasting},
  edition   = {2nd},
  publisher = {Springer},
  year      = {2002}
}

@article{Paolillo:2023ssv,
    author = "Paolillo, M. and others",
    title = "{The universal shape of the X-ray variability power spectrum of AGN up to z {\ensuremath{\sim}} 3}",
    eprint = "2302.08524",
    archivePrefix = "arXiv",
    primaryClass = "astro-ph.HE",
    doi = "10.1051/0004-6361/202245291",
    journal = {\aap},
    volume = "673",
    pages = "A68",
    year = "2023"
}

@ARTICLE{2025MNRAS.541.1851B,
       author = {{Bright}, Joe S. and {Fender}, Rob and {Russell}, David M. and {Motta}, Sara E. and {Man}, Ethan and {van den Eijnden}, Jakob and {Alabarta}, Kevin and {Crook-Mansour}, Justine and {Baglio}, Maria C. and {Green}, David A. and {Heywood}, Ian and {Lewis}, Fraser and {Saikia}, Payaswini and {Scott}, Paul F. and {Titterington}, David J.},
        title = "{The accretion{\textendash}ejection connection in the black hole X-ray binary MAXI J1820+070}",
      journal = {\mnras},
         year = 2025,
        month = aug,
       volume = {541},
       number = {2},
        pages = {1851-1865},
          doi = {10.1093/mnras/staf1098},
archivePrefix = {arXiv},
       eprint = {2507.11303},
 primaryClass = {astro-ph.HE},
       adsurl = {https://ui.adsabs.harvard.edu/abs/2025MNRAS.541.1851B}
}

@ARTICLE{tramacere2009,
  author  = {Tramacere, A. and Giommi, P. and Perri, M. and
             Verrecchia, F. and Tosti, G.},
  title   = {{Swift} observations of the very intense flaring activity
             of {Mrk~421} during 2006.~{I.} Multiwavelength analysis
             of four intense flares},
  journal = {\aap},
  year    = {2009},
  volume  = {501},
  pages   = {879--898},
  doi     = {10.1051/0004-6361/200810865}
}

@ARTICLE{kirk1998,
  author  = {Kirk, J.~G. and Rieger, F.~M. and Mastichiadis, A.},
  title   = {Particle acceleration and synchrotron emission in blazar jets},
  journal = {\aap},
  year    = {1998},
  volume  = {333},
  pages   = {452--458},
  archivePrefix = {arXiv},
  eprint  = {astro-ph/9801265}
}

@ARTICLE{li2000,
  author  = {Li, H. and Kusunose, M.},
  title   = {Temporal and Spectral Variabilities of High-Energy Emission
             from Blazars Using Synchrotron Self-{C}ompton Models},
  journal = {\apj},
  year    = {2000},
  volume  = {536},
  pages   = {729--742},
  doi     = {10.1086/308971},
  archivePrefix = {arXiv},
  eprint  = {astro-ph/0002134}
}

@ARTICLE{fossati2000a,
  author  = {Fossati, G. and Celotti, A. and Chiaberge, M. and
             Zhang, Y.~H. and Chiappetti, L. and Ghisellini, G. and
             Maraschi, L. and Tavecchio, F. and Pian, E. and Treves, A.},
  title   = {{X}-ray Emission of {M}kn~421: New Clues From Its Spectral
             Evolution. {I}. Temporal Analysis},
  journal = {\apj},
  year    = {2000},
  volume  = {541},
  pages   = {153--165},
  doi     = {10.1086/309411},
  archivePrefix = {arXiv},
  eprint  = {astro-ph/0005066}
}

@article{Kapanadze:2020qpy,
    author = "Kapanadze, B. and Gurchumelia, A. and Dorner, D. and Vercellone, S. and Romano, P. and Hughes, P. and Aller, M. and Aller, H. and Kharshiladze, O.",
    title = "{Swift Observations of Mrk 421 in Selected Epochs. III. Extreme X-Ray Timing/Spectral Properties and Multiwavelength Lognormality during 2015 December{\textendash}2018 April}",
    eprint = "2004.00676",
    archivePrefix = "arXiv",
    primaryClass = "astro-ph.GA",
    doi = "10.3847/1538-4365/ab6322",
    journal = "Astrophys. J. Suppl.",
    volume = "247",
    number = "1",
    pages = "27",
    year = "2020"
}

@article{Kapanadze:2024nwg,
    author = "Kapanadze, B. and Gurchumelia, A. and Aller, M.",
    title = "{Swift Observations of Mrk 421 in Selected Epochs. IV. Physical Implications of X-Ray Flaring Activity and Features of Relativistic Magnetic Reconnection in 2018 April{\textendash}2023 December}",
    doi = "10.3847/1538-4365/ad7d0c",
    journal = "Astrophys. J. Suppl.",
    volume = "275",
    number = "2",
    pages = "23",
    year = "2024"
}

@article{Levy:2024eiq,
    author = "Levy, C. and Sol, H. and Bolmont, J.",
    title = "{Separating source-intrinsic and Lorentz invariance violation induced delays in the very high-energy emission of blazar flares}",
    eprint = "2406.01182",
    archivePrefix = "arXiv",
    primaryClass = "astro-ph.HE",
    doi = "10.1051/0004-6361/202450140",
    journal = "Astron. Astrophys.",
    volume = "689",
    pages = "A136",
    year = "2024"
}

@article{Perennes:2019sjx,
    author = "Perennes, C. and Sol, H. and Bolmont, J.",
    title = "{Modeling spectral lags in active galactic nucleus flares in the context of Lorentz invariance violation searches}",
    eprint = "1911.10377",
    archivePrefix = "arXiv",
    primaryClass = "astro-ph.HE",
    doi = "10.1051/0004-6361/201936430",
    journal = "Astron. Astrophys.",
    volume = "633",
    pages = "A143",
    year = "2020"
}

@article{MAGIC:2024mjz,
    author = "Abe, K. and others",
    collaboration = "MAGIC",
    title = "{Insights from the first flaring activity of a high synchrotron peaked blazar with X-ray polarization and VHE gamma rays}",
    eprint = "2410.23140",
    archivePrefix = "arXiv",
    primaryClass = "astro-ph.HE",
    doi = "10.1051/0004-6361/202452785",
    journal = "Astron. Astrophys.",
    volume = "695",
    pages = "A217",
    year = "2025"
}

@article{MAGIC:2025omm,
    author = "Abe, K. and others",
    collaboration = "MAGIC",
    title = "{Time-dependent Modeling of the Subhour Spectral Evolution during the 2013 Outburst of Mrk 421}",
    eprint = "2509.08686",
    archivePrefix = "arXiv",
    primaryClass = "astro-ph.HE",
    doi = "10.3847/1538-4357/ae25f0",
    journal = "Astrophys. J.",
    volume = "998",
    number = "1",
    pages = "6",
    year = "2026"
}

%%%%%%%%%%%%%%%%%%%%%%%%%%
\appendix
\nolinenumbers
%%%%%%%%%%%%%%%%%%%%%%%%%%
\section{Relationship between $f_\mathrm{cl}$ and $d_\mathrm{cl}$}
\label{app:closure}

\begin{figure}
  \centering
  \includegraphics[width=\columnwidth]{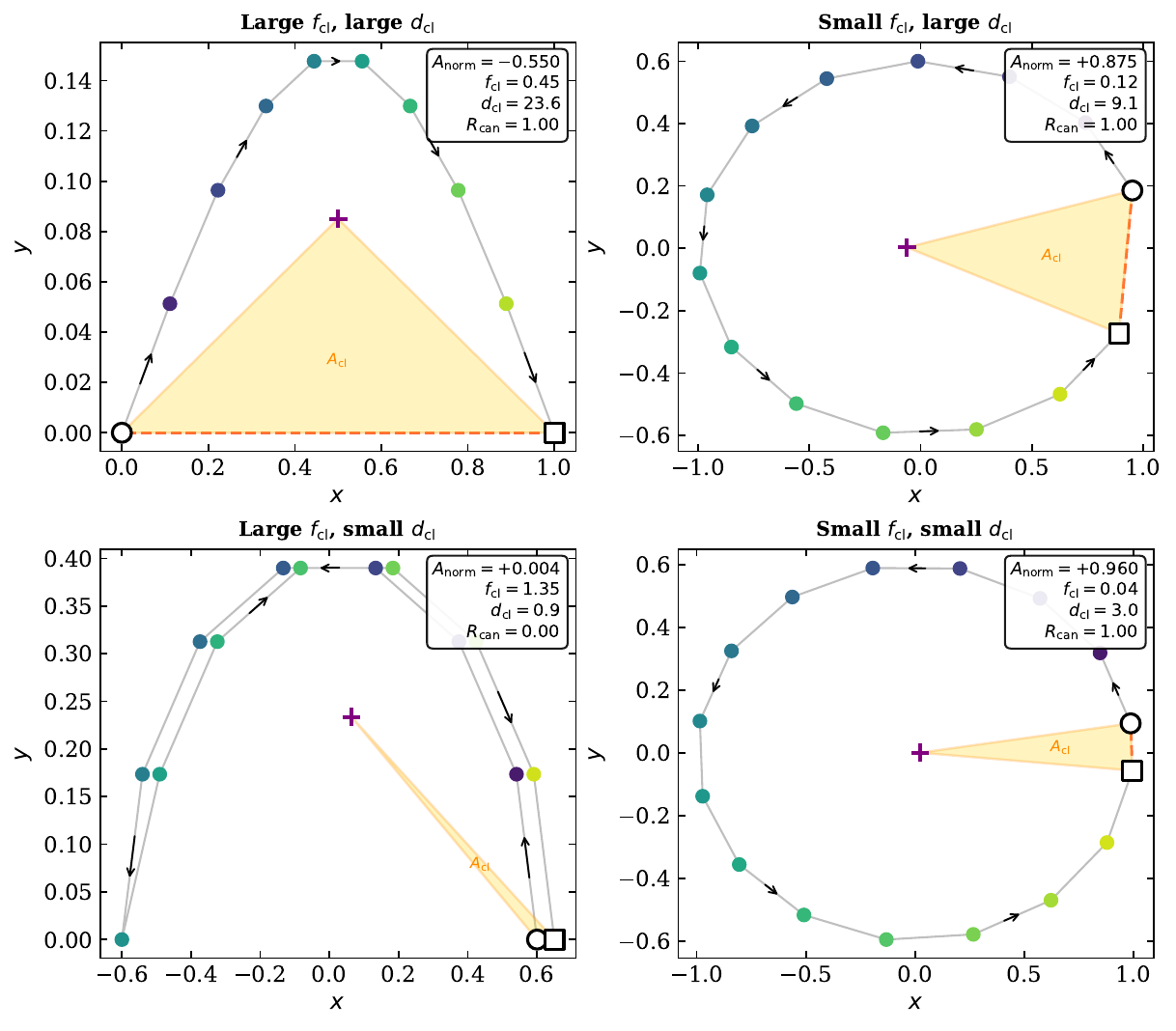}
  \caption{Four illustrative trajectories demonstrating the relationship between $f_\mathrm{cl}$ and $d_\mathrm{cl}$. Open circles mark the start point; open squares mark the end point; the dashed red line is the artificial closing segment. Measured statistics are given in the top-right corner of each panel. \emph{Top left:} large $f_\mathrm{cl}$, large $d_\mathrm{cl}$---closing segment dominates. \emph{Top right:} small $f_\mathrm{cl}$, large $d_\mathrm{cl}$---large loop, incomplete cycle. \emph{Bottom left:} large $f_\mathrm{cl}$, small $d_\mathrm{cl}$---degenerate near-cancellation; also indicated by $R_\mathrm{can}$. \emph{Bottom right:} small $f_\mathrm{cl}$, small $d_\mathrm{cl}$---complete loop, ideal case.}
  \label{fig:closure_examples}
\end{figure}

The closure fraction $f_\mathrm{cl}$ and the normalised closure distance $d_\mathrm{cl}$ both describe the geometry of the trajectory endpoints, but they measure different things and are largely non-redundant. Fig.~\ref{fig:closure_examples} illustrates the four combinations of large and small values of each diagnostic using constructed trajectories.

The top-left panel shows the case that most naturally produces large $f_\mathrm{cl}$. A nearly straight open path where the endpoints are far apart ($d_\mathrm{cl}$ large) and the closing segment cuts across the interior, contributing most of the total enclosed area. 
The top-right panel shows that large $d_\mathrm{cl}$ does not imply large $f_\mathrm{cl}$. A well-developed loop where the source was observed through approximately three-quarters of its cycle has widely separated endpoints, yet the open path already encloses most of the area. The closing segment clips only a small corner, so $f_\mathrm{cl}$ is small and $A_\mathrm{open}$ is reliable. $d_\mathrm{cl}$ correctly flags that the cycle is incomplete, but this does not invalidate the measurement. 
The bottom-left panel shows the only way to achieve large $f_\mathrm{cl}$ with small $d_\mathrm{cl}$. A figure-eight or strongly cancelling trajectory where $A_\mathrm{tot} \approx 0$, causing $f_\mathrm{cl} = |A_\mathrm{closure}|/|A_\mathrm{tot}|$ to blow up even though both $A_\mathrm{open}$ and $A_\mathrm{closure}$ are individually small. This is a degenerate case in which $R_\mathrm{can}$ is also small. 
The bottom-right panel shows the ideal case: a nearly complete loop with endpoints close together and $f_\mathrm{cl}$ small. 

In summary, $f_\mathrm{cl}$ and $d_\mathrm{cl}$ are largely non-redundant diagnostics that capture different aspects of trajectory completeness. We verified through simulations of null trajectories that neither quantity has sufficient predictive power for $|A_\mathrm{norm}|$ to serve as a reliable hard detection criterion: $f_\mathrm{cl}$ has very wide scatter at any given value, while $d_\mathrm{cl}$ shows essentially no correlation with $|A_\mathrm{norm}|$ across the null distribution. Both are, therefore, reported as diagnostics alongside the primary statistics; large values indicate that the observed trajectory did not cover a complete cycle and should be noted when reporting results.

%This figure is produced by \texttt{appendix\_closure\_figure.py}.} 

%%%%%%%%%%%%%%%%%%%%%%%%%%
\section{Self-intersecting trajectories and the $|A_\mathrm{norm}|>1$
exception}
\label{app:winding}

\begin{figure}
  \centering
  \includegraphics[width=\columnwidth]{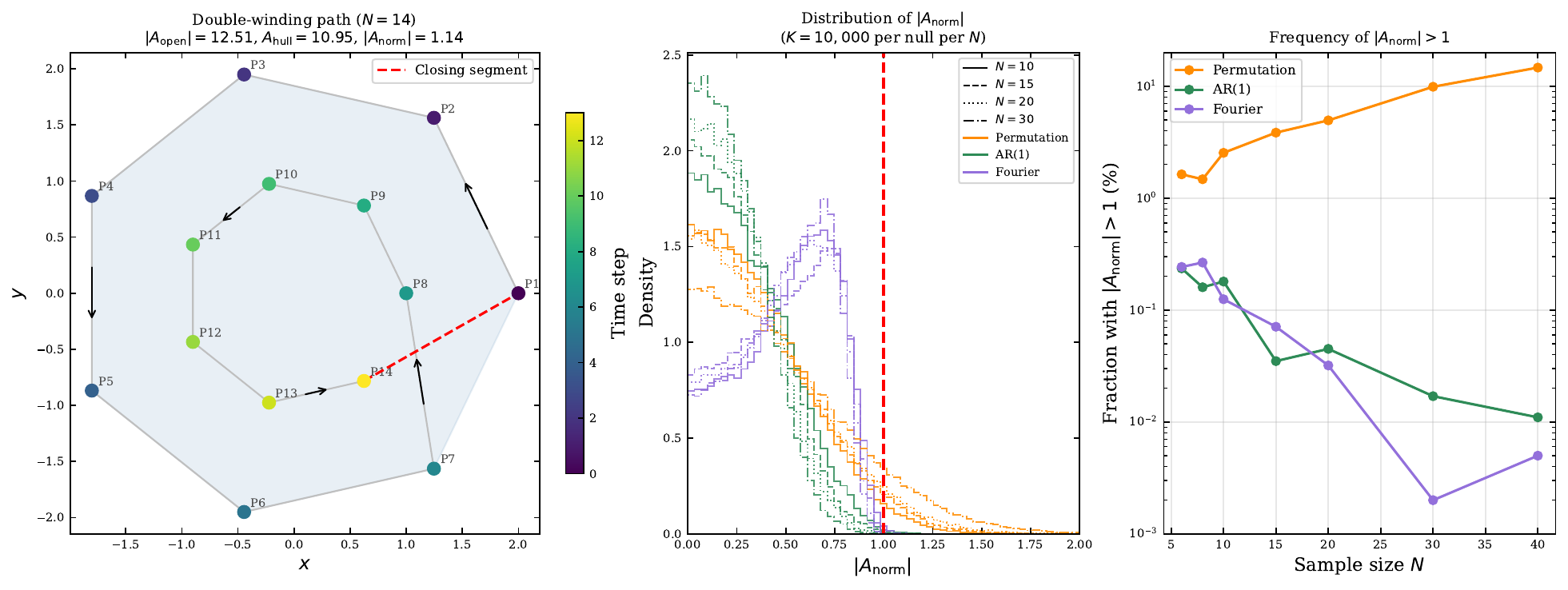}
  \caption{\emph{Left:} explicit double-winding trajectory ($N=14$) for which $|A_\mathrm{norm}| = 1.14 > 1$. The path traces a 7-point outer circle (radius 2) followed by a 7-point inner circle (radius 1) in the same CCW direction. The shaded region shows the convex hull. \emph{Centre:} distribution of $|A_\mathrm{norm}|$ for all three null models at four sample sizes ($K = 50{,}000$ realisations each); the dashed line marks $|A_\mathrm{norm}| = 1$. \emph{Right:} fraction of null realisations with $|A_\mathrm{norm}| > 1$ as a function of $N$ for all three null models.} 
  \label{fig:winding}
\end{figure}

For a simple (non-self-intersecting) closed polygon, the shoelace area cannot exceed the convex hull area, so $|A_\mathrm{norm}| \leq 1$ is guaranteed. For self-intersecting paths the shoelace formula counts each enclosed region with its winding number, meaning a path that winds twice around a region in the same direction counts that region's area twice. A simple explicit example (see the left panel in Fig.~\ref{fig:winding}) is a path that traces a large circle followed by a concentric smaller circle in the same direction: the inner disc is counted by both loops, giving $|A_\mathrm{norm}| > 1$.

In simulations of $K = 50{,}000$ null realisations using the signal~(a) trajectory as the base (see Fig.~\ref{fig:winding}), $|A_\mathrm{norm}| > 1$ occurs in the fraction of realisations shown in the right panel of Fig.~\ref{fig:winding} for $N = 6$--$40$. The permutation null produces the most such cases by a wide margin: $\approx 1.2\%$ at $N=6$, rising to $\approx 18\%$ at $N=40$. AR(1) and Fourier are both below $0.6\%$ at all $N$ and decrease with increasing $N$. In all three null models, the trajectories with $|A_\mathrm{norm}| > 1$ have large $R_\mathrm{can}$ (mean $\approx 0.96$ for AR(1), $\approx 0.91$ for permutation, $\approx 1.00$ for Fourier), because the winding passes proceed in the same direction with no cancellation between oppositely-oriented sub-loops. Their $|A_\mathrm{norm}|$ values exceed 1.0 by a small margin for AR(1) and Fourier (maximum $\approx 1.2$ and $\approx 1.1$ respectively), while for permutation the excess can be larger at high $N$ (maximum $\approx 2.7$ at $N=30$). Despite the high permutation fraction at large $N$, the practical effect on reported $p$-values is negligible at the sample sizes typical of the present application ($N \lesssim 20$).

%%%%%%%%%%%%%%%%%%%%%%%%%%
\section{Comparison of $A_\mathrm{abs}$ and $A_\mathrm{rms}$}
\label{app:arms}

Both $A_\mathrm{abs}$ and $A_\mathrm{rms}$ measure the total rotational content irrespective of orientation, but they weight the individual triangle areas differently. $A_\mathrm{abs}$ sums contributions linearly, while $A_\mathrm{rms}$ squares each contribution, and thus upweights large contributions relative to small ones.

For a trajectory with $N$ points, the $N-1$ signed triangle areas $a_i$ satisfy the inequality
\begin{equation}
    \frac{A_\mathrm{abs}}{\sqrt{N-1}} \leq A_\mathrm{rms} \leq A_\mathrm{abs}. 
\end{equation}
The upper bound $A_\mathrm{rms} = A_\mathrm{abs}$ is reached when only a single triangle is non-zero; the lower bound $A_\mathrm{rms} = A_\mathrm{abs}/\sqrt{N-1}$ is reached when all $N-1$ triangles have equal magnitude. All trajectories must therefore fall within the band bounded by these two lines. 

\begin{figure}
  \centering
  \includegraphics[width=\columnwidth]{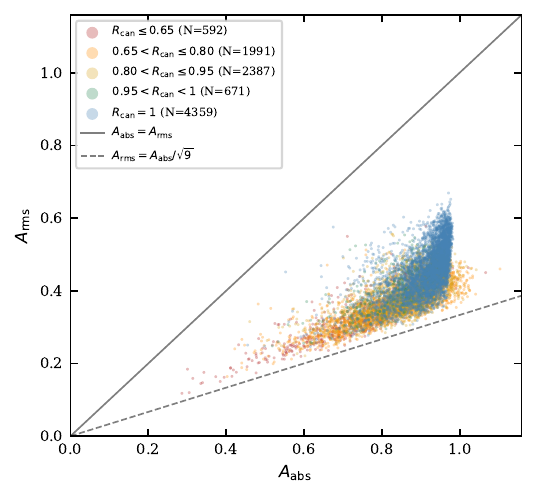}
  \caption{Normalised $A_\mathrm{rms}$ vs.\ normalised $A_\mathrm{abs}$ for a sample of $10^4$ loops ($N=10$ points each) produced with the toy model. Points are colour-coded by $R_\mathrm{can}$. The solid line marks $A_\mathrm{rms} = A_\mathrm{abs}$, and the dashed one $A_\mathrm{rms} = A_\mathrm{abs}/3$.}
  \label{fig:arms}
\end{figure}

Fig.~\ref{fig:arms} shows normalised $A_\mathrm{rms}$ vs. normalised $A_\mathrm{abs}$ for $10^4$ trajectories generated with the toy model (Sect.~\ref{sec:model}), colour-coded according to the $R_\mathrm{can}$ range. The upper and lower bounds are shown with solid and dashed lines, respectively. 

We adopt $A_\mathrm{abs}$ as the primary magnitude diagnostic because: (i) it has a direct geometric interpretation as the total unsigned area swept by the trajectory; (ii) it scales linearly with signal amplitude; (iii) its considerably larger dynamic range makes it more sensitive to differences in trajectory morphology; and (iv) it does not artificially suppress the cumulative contribution of measurement noise, keeping the statistic honest about the total rotational content of the observed trajectory.

%%%%%%%%%%%%%%%%%%%%%%%%%%
\section{Investigating corrected approaches for open loops}
\label{app:corrections}

Here, we discuss different options for correcting or reweighting $|A_\mathrm{norm}|$ in case of open loops.

\paragraph{Attempted correction via endpoint conditioning.}
One approach to this problem is to condition the null distributions on the observed endpoint separation: accept only surrogate realisations whose endpoint separation $\Delta_\mathrm{null}$ falls within a tolerance $\tau_\Delta$ of $\Delta_\mathrm{obs}$, where $\tau_\Delta$ is set to one standard deviation of the propagated uncertainty on $\Delta_\mathrm{obs}$ (Eq.~\ref{eq:sigmadelta}). For trajectories that are nearly closed, this conditioning has a negligible effect since the null distributions already concentrate near $\Delta = 0$.

However, for open trajectories, using numerical tests, we found that the distribution of endpoint separations $\Delta_\mathrm{null}$ produced by the nulls rarely overlaps with the acceptance window $[\Delta_\mathrm{obs} - \sigma_\Delta,\, \Delta_\mathrm{obs} + \sigma_\Delta]$ (see Eqs.~\ref{eq:Delta_obs},~\ref{eq:sigmadelta}).

\paragraph{Attempted correction via closure-fraction scaling.}
A second approach replaces $|A_\mathrm{norm}|$ as the test statistic with $T = |A_\mathrm{norm}|/(1 - f_\mathrm{cl})$, applied to both the observed trajectory and each null surrogate. The motivation follows from the identity
\begin{equation}
  \frac{|A_\mathrm{norm}^\mathrm{obs}|}{|A_\mathrm{norm}^\mathrm{null}|}
  = \frac{1 - f_\mathrm{cl}^\mathrm{obs}}{1 - f_\mathrm{cl}^\mathrm{null}},
  \label{eq:fcl_scaling}
\end{equation}
which states that comparing $T^\mathrm{obs}$ to $T^\mathrm{null}$ places the observed and null trajectories on equal footing with respect to their closure fractions.

The approach breaks down when $A_\mathrm{open}$ and $A_\mathrm{closure}$ have opposite signs, giving $f_\mathrm{cl} > 1$ and making $T$ undefined. This occurs in 10--20\% of permutation null realisations and a smaller but non-negligible fraction of AR(1) and Fourier realisations, depending on the trajectory. No principled definition of $T$ exists for these cases within the current framework. Excluding them would bias the $p$-value by reducing the effective sample size. 

We therefore do not apply endpoint conditioning in the default analysis. Diagnostics $f_\mathrm{cl}$ and $d_\mathrm{cl}$ serve as qualitative indicators that the detection power may be reduced for open trajectories,
and analysers are advised to interpret the reported $p$-values in that context.

\end{document}